\documentclass[12pt,leqno]{article}
\title{Generalized thermodynamics and kinetic equations: Boltzmann, Landau, Kramers and Smoluchowski}

\usepackage{amsthm,amsmath,amssymb,a4wide,psfig,epsf}
\def\mb#1{\setbox0=\hbox{$#1$}\kern-.025em\copy0\kern-\wd0
\kern-0.05em\copy0\kern-\wd0\kern-.025em\raise.0233em\box0}

\begin{document}

\author{Pierre-Henri Chavanis}
\maketitle
\begin{center}
Laboratoire de Physique Th\'eorique, Universit\'e Paul Sabatier,\\
118, route de Narbonne, 31062 Toulouse, France\\ E-mail: {\it
chavanis{@}irsamc.ups-tlse.fr }

%\date{}
\vspace{0.5cm}
\end{center}

\begin{abstract}

We propose a formal extension of thermodynamics and kinetic
theories to a larger class of entropy functionals. Kinetic
equations associated to Boltzmann, Fermi, Bose and Tsallis
entropies are recovered as a special case. This formalism first
provides a unifying description of classical and quantum kinetic
theories. On the other hand, a generalized thermodynamical
framework is justified to describe complex systems exhibiting
anomalous diffusion. Finally, a notion of generalized
thermodynamics emerges in the context of the the violent
relaxation of collisionless stellar systems and two-dimensional
vortices due to the existence of Casimir invariants and incomplete
relaxation. A thermodynamical analogy can also be developed to
analyze the nonlinear dynamical stability of stationary solutions
of the Vlasov and 2D Euler-Poisson systems. On general grounds, we
suggest that generalized entropies arise due to the existence of
``hidden constraints'' that modify the form of entropy that we would
naively expect. Generalized kinetic equations are therefore 
``effective'' equations that are introduced heuristically to describe
complex systems. 

\end{abstract}

\section{Introduction}
\label{intro}

Standard kinetic equations satisfy two fundamental properties linked
to the first and second principles of thermodynamics: the conservation
of energy (and mass) and the increase of entropy (H-theorem). These
properties are shared in particular by the Boltzmann and by the Landau
equations which are at the basis of the kinetic theory of dilute gases
and neutral plasmas \cite{balescu}. When the system is in contact with
a thermostat, instead of being isolated, the conservation of energy
and the increase of entropy (microcanonical description) are replaced
by the decrease of free energy $F=E-TS$ at fixed temperature
(canonical description). This is the proper description of Brownian
motion that is analyzed in terms of stochastic processes and
Fokker-Planck equations (Kramers, Smoluchowski,...) \cite{risken}. In
the standard framework, the functional increasing monotonically with
time is the Boltzmann entropy $S_B[f]=-\int f \ln f d^{3}{\bf
r}d^{3}{\bf v}$ or the Boltzmann free energy $J_B[f]=S_{B}[f]-\beta E[f]$.

In a recent paper \cite{gt}, we have proposed to develop a
generalized thermodynamical formalism for a larger class of
functionals that we called {\it generalized entropies}. They can
be written $S[f]=-\int C(f)d^{3}{\bf r}d^{3}{\bf v}$ where $C(f)$
is a convex function, i.e. $C''(f)>0$. Boltzmann, Fermi, Bose, and
Tsallis entropies are particular functionals of the above form. On
general grounds, what we mean by generalized thermodynamics is the
extension of the usual variational principle of classical
thermodynamics (maximization of Boltzmann entropy $S_{B}[f]$ at
fixed mass $M$ and energy $E$) to a larger class of functionals.
This variational principle arises in many problems of physics (or
biology, economy,...) for various reasons that do not have
necessarily a direct relation to thermodynamics. In many cases, it
is relevant to develop a {\it thermodynamical analogy} and to use
the same vocabulary as in standard thermodynamics. This allows one
to transpose directly the standard methods developed in ordinary
thermodynamics to a new context.

In \cite{gt} we have proposed a generalized class of Fokker-Planck
equations associated to this generalized thermodynamical
framework. This formalism is interesting to develop because it
leads to a unified description of known kinetic theories
(classical and quantum) and it also generates new types of kinetic
equations. In that respect, it can be of interest in applied
mathematics and theoretical physics. This formalism can also have
important physical applications. For example, these generalized
Fokker-Planck equations can account for a process of {\it
anomalous diffusion} in complex media. They arise naturally from
ordinary Fokker-Planck equations by assuming that the diffusion
coefficient is a function of the density. Generalized
thermodynamics can also be relevant for the violent relaxation of
stellar systems and two-dimensional (2D) vortices. In that
context, generalized entropies (also called $H$-functions) emerge
due to the existence of fine-grained constraints (Casimir
invariants) that modify the form of entropy that we would naively
expect. Generalized Fokker-Planck equations can provide a simple
small-scale parametrization of turbulence (mixing) in stellar
dynamics and 2D hydrodynamics. They can also serve as powerful
{\it numerical algorithms} to compute arbitrary nonlinearly dynamically
stable solutions of the 2D Euler-Poisson or Vlasov-Poisson
systems. Indeed, the condition of nonlinear dynamical stability
can be put in a form analogous to a condition of generalized
thermodynamical stability \cite{ipser,tremaine,ellis,grand}. This
is a striking illustration of the {\it thermodynamical analogy}
mentioned above. Since the notion of generalized thermodynamics
can have different interpretations, it is relevant to work at a
general level and develop a formalism without explicit reference
to a precise context. Then, a justification of this generalization
and a physical interpretation of the results must be given in each
case.

In an early work \cite{csr}, we observed that classical and
quantum Fokker-Planck equations can be obtained from a
phenomenological Maximum Entropy Production Principle (MEPP)
\cite{rsmepp} by maximizing the rate of entropy (resp. free
energy) production at fixed mass and energy (resp. temperature).
This variational approach, closely related to the linear
thermodynamics of Onsager, is the most natural extension of the
equilibrium thermodynamical principle in which we maximize entropy
(resp. free energy) at fixed mass and energy (resp. temperature).
In \cite{gt}, we generalized this principle to a larger class of
entropy functionals and obtained generalized Fokker-Planck
equations. We also ``guessed'' a form of generalized Landau
equation consistent to the generalized Kramers equation obtained
with the MEPP. Recently, we found that a similar generalization of
kinetic theory was attempted by Kaniadakis \cite{kaniadakis} from
a different point of view. His approach consists in generalizing
the assumptions that are made at the start to derive the Boltzmann
and the Fokker-Planck equations. This amounts to modifying the
form of the transition probabilities that arise in the dynamical
process. Such generalized transition probabilities are relevant
for quantum particles (fermions and bosons) and possibly, also, in
the physics of complex media. The MEPP approach, on the other
hand, is purely thermodynamical and exploits at best the first and
second principles of thermodynamics (possibly extended to
generalized functionals) in a viewpoint reminiscent of Jaynes'
ideas.

In this paper, we show that the two approaches lead to equivalent
kinetic equations. In Sec. \ref{sec_gk}, we use Kaniadakis
approach to derive the generalized Landau equation from the
generalized Boltzmann equation in a {\it weak deflexion
approximation}. In Sec. \ref{sec_vr}, we show that the generalized
Landau equation can also be obtained by coarse-graining the Vlasov
equation in the context of the violent relaxation of stellar
systems. In Sec. \ref{sec_prop}, we establish the main properties
of the generalized Landau equation (conservation laws, generalized
H-theorem,...). In Sec. \ref{sec_tb}, we derive the generalized
Kramers equation from the generalized Landau equation in a test
particle approach and a {\it thermal bath approximation}. The
generalized Smoluchowski equation is in turn derived from the
generalized Kramers equation in a {\it hydrodynamical limit}. In
Secs. \ref{sec_diff} and \ref{sec_exp}, we determine explicit
expressions of the diffusion coefficient for Boltzmann, Fermi and
Tsallis distributions. In Sec. \ref{sec_escape}, we use the
generalized Kramers equation to derive generalized truncated
distribution functions accounting for an escape of particles above
a limit energy.

\section{Generalized kinetic equations}
\label{sec_gk}

\subsection{The generalized Boltzmann equation}
\label{sec_bol}

The ordinary Boltzmann equation can be written in the form
\begin{equation}
\label{bol1}
{d f\over d t}=\int d^{3}{\bf v}_{1}\ d\Omega \ w({\bf v},{\bf v}_{1};{\bf v}',{\bf v}_{1}')\biggl\lbrace f({\bf v}')f({\bf v}'_{1})-f({\bf v})f({\bf v}_{1})\biggr\rbrace,
\end{equation}
where $d\Omega$ is the element of solid angle and $w({\bf v},{\bf
v}_{1};{\bf v}',{\bf v}_{1}')$ is the density probability of a
collision transforming the velocities ${\bf v}$,${\bf v}_{1}$ in ${\bf
v}'$,${\bf v}'_{1}$ or the converse (the abbreviations $f({\bf v}')$,
$f({\bf v}'_{1})$, $f({\bf v})$, $f({\bf v}_{1})$ stand for for
$f({\bf r},{\bf v}',t)$, $f({\bf r},{\bf v}'_{1},t)$, $f({\bf r},{\bf
v},t)$, $f({\bf r},{\bf v}',t)$). The material derivative is
$d/dt=\partial/\partial t+{\bf v}\partial/\partial {\bf r}+{\bf
F}\partial/\partial {\bf v}$ where ${\bf F}=-\nabla\Phi$ is a
mean-field force acting on the particles. The Boltzmann equation
conserves the mass
\begin{equation}
\label{bol1a}
M=\int \rho d^{3}{\bf r},
\end{equation}
the energy
\begin{equation}
\label{bol1b}
E=\int f {v^{2}\over 2}d^{3}{\bf r}d^{3}{\bf v}+{1\over 2}\int \rho\Phi d^{3}{\bf r},
\end{equation}
the angular momentum
\begin{equation}
\label{bol1c}
L=\int f {\bf r}{\times} {\bf v} d^{3}{\bf r}d^{3}{\bf v},
\end{equation}
and the impulse
\begin{equation}
\label{bol1d}
{\bf P}=\int f {\bf v} d^{3}{\bf r}d^{3}{\bf v},
\end{equation}
where $\rho({\bf r},t)$ is the spatial density. In addition, the
Boltzmann entropy
\begin{equation}
\label{bol1e}
S=-\int f \ln f  d^{3}{\bf r}d^{3}{\bf v},
\end{equation}
satisfies a $H$-theorem, i.e. $\dot S\ge 0$ with $\dot S=0$ if, and only if, the distribution $f({\bf r},{\bf v})$ is the Boltzmann distribution
\begin{equation}
\label{bol1f}
f_{eq}({\bf r},{\bf v})=A e^{-\beta\epsilon'},
\end{equation}
where $\epsilon'={v^{2}\over 2}+\Phi+{\mb\Omega}\cdot ({\bf r}{\times} {\bf v})+{\bf U}\cdot {\bf v}$ is the energy of a particle by unit of mass.

Recently, Kaniadakis
\cite{kaniadakis} has proposed the generalization
\begin{equation}
\label{bol2}
{d f\over dt}=\int d^{3}{\bf v}_{1}\ d\Omega \ w({\bf v},{\bf v}_{1};{\bf v}',{\bf v}_{1}')\biggl\lbrace a(f')b(f)a(f_{1}')b(f_{1})-a(f)b(f')a(f_{1})b(f_{1}')\biggr\rbrace,
\end{equation}
where the functions $a(f)$ and $b(f)$ are somewhat arbitrary (we
have noted $f=f({\bf v})$, $f_{1}=f({\bf v}_{1})$, $f'=f({\bf
v}')$ and $f_{1}'=f({\bf v}'_{1})$).  This generalization
encompasses the case of quantum particles (fermions and bosons)
with exclusion or inclusion principles. This generalization could
also be relevant in the case of complex systems for which the
transition probabilities are not simply the product of distribution
functions. This can happen when we are not in the strict conditions
of validity of the ordinary Boltzmann equation. It is also possible that
Eq. (\ref{bol2}) is just an {\it effective} kinetic equation
accounting for ``hidden constraints'' in complex media.

In the following, we shall consider the situation in which the
potential of interaction between particles is Coulombian (or
Newtonian). In that case, each encounter provokes a weak deflexion of
the particles trajectory and it is of order to consider the weak
deflexion limit of the Boltzmann equation.  Classically, this leads
to the so-called Landau equation which forms the basis of the kinetic
theory of neutral plasmas \cite{balescu} and stellar systems
\cite{bt}.  Our aim is this section is to derive a generalized Landau
equation from the generalized Boltzmann equation proposed by
Kaniadakis. This generalization is essentially formal. In the
following, we follow the classical derivation of the Landau equation
reported in the monograph of Balescu \cite{balescu}. Therefore, we
shall omit the calculations that are identical to the classical case
and refer to \cite{balescu} for more details.

\subsection{The weak deflexion approximation}
\label{sec_wd}

First, it is convenient to write the velocities of the particles
before and after the collision as
\begin{equation}
\label{wd1}
{\bf v}'={\bf v}+{\bf \Delta},
\end{equation}
\begin{equation}
\label{wd2}
{\bf v}'_{1}={\bf v}_{1}-{\bf \Delta},
\end{equation}
where ${\bf\Delta}$ is the velocity deviation. We can now express
the probability of a collision in terms of new variables as \cite{balescu}:
\begin{equation}
\label{wd3a}
w({\bf v},{\bf v}_{1};{\bf v}',{\bf v}_{1}')\rightarrow w({\bf v}+{{\bf\Delta}\over 2},{\bf v}_{1}-{{\bf\Delta}\over 2};{\bf\Delta}).
\end{equation}
The generalized Boltzmann equation can thus be rewritten
\begin{eqnarray}
\label{wd4}
{df\over  dt}=\int d^{3}{\bf v}_{1}\ d\Omega\ w({\bf v}+{{\bf\Delta}\over 2},
{\bf v}_{1}-{{\bf\Delta}\over 2};{\bf\Delta})
\biggl\lbrace a\lbrack f({\bf v}+{\bf \Delta})\rbrack b\lbrack f({\bf v})\rbrack a\lbrack f({\bf v}_{1}-{\bf \Delta})\rbrack b\lbrack f({\bf v}_{1})\rbrack \nonumber\\
-a\lbrack f({\bf v})\rbrack b\lbrack f({\bf v}+{\bf \Delta})\rbrack  a\lbrack f({\bf v}_{1})\rbrack b\lbrack f({\bf v}_{1}-{\bf \Delta})\rbrack\biggr\rbrace.
\end{eqnarray}
In the weak deflexion limit $|{\bf \Delta}|\ll |{\bf v}|, |{\bf
v_{1}}|$, we can expand the r.h.s. of Eq. (\ref{wd4}) in Taylor series.
Using
\begin{equation}
\label{wd5}
 w({\bf v}+{{\bf\Delta}\over 2},{\bf v}_{1}-{{\bf\Delta}\over 2};{\bf\Delta})\simeq  w({\bf v},{\bf v}_{1};{\bf\Delta})+{1\over 2}\Delta^{\mu}\biggl\lbrace {\partial w({\bf v},{\bf v_{1}};{\bf\Delta})\over\partial v^{\mu}}- {\partial w({\bf v},{\bf v_{1}};{\bf\Delta})\over\partial v_{1}^{\mu}}\biggr\rbrace+...,
\end{equation}
\begin{equation}
\label{wd6}
f({\bf v}+{\bf\Delta})=f({\bf v})+\Delta^{\mu}{\partial f({\bf v})\over\partial v^{\mu}}+{1\over 2}\Delta^{\mu}\Delta^{\nu}{\partial^{2}f({\bf v})\over\partial v^{\mu}\partial v^{\nu}}+...,
\end{equation}
we get
\begin{eqnarray}
\label{wd7}
{df\over dt}={1\over 2}\int d^{3}{\bf v}_{1}\ d\Omega \Delta^{\mu}\Delta^{\nu}\biggl\lbrace  w({\bf v},{\bf v}_{1};{\bf\Delta})\biggl\lbrack ab(b_{1}a'_{1}-a_{1}b'_{1}){\partial^{2}f_{1}\over\partial v_{1}^{\mu}\partial v_{1}^{\nu}}+ab(b_{1}a''-a_{1}b''){\partial f_{1}\over\partial v_{1}^{\mu}}{\partial f_{1}\over\partial v_{1}^{\nu}}\nonumber\\
+a_{1}b_{1}(ba'-b'a){\partial^{2}f\over\partial v^{\mu}\partial v^{\nu}}
-2(b b_{1} a' a'_{1}-a a_{1} b' b'_{1}){\partial f\over\partial v^{\mu}}{\partial f_{1}\over\partial v_{1}^{\nu}}+a_{1}b_{1}(ba''-ab'') {\partial f\over\partial v^{\mu}}{\partial f\over\partial v^{\nu}}\biggr\rbrack\nonumber\\
+{\partial w\over\partial v^{\mu}}\biggl\lbrack a_{1}b_{1}(ba'-b'a){\partial f\over\partial v^{\nu}}-ab(b_{1}a'_{1}-a_{1}b'_{1}){\partial f_{1}\over\partial v_{1}^{\nu}}\biggr\rbrack\nonumber\\
-{\partial w\over\partial v_{1}^{\mu}}\biggl\lbrack a_{1}b_{1}(ba'-b'a){\partial f\over\partial v^{\nu}}-ab(b_{1}a'_{1}-a_{1}b'_{1}){\partial f_{1}\over\partial v_{1}^{\nu}}\biggr\rbrack\biggr\rbrace,
\end{eqnarray}
where $a=a\lbrack f({\bf v})\rbrack$, $a_{1}=a\lbrack f({\bf
v}_{1})\rbrack$, $a'=a'\lbrack f({\bf v})\rbrack$,
$a'_{1}=a'\lbrack f({\bf v}_{1})\rbrack$ etc... Integrating the
last term by parts, the foregoing expression simplifies in
\begin{eqnarray}
\label{wd8}
{df\over dt}={1\over 2}\int d^{3}{\bf v}_{1}\ d\Omega \Delta^{\mu}\Delta^{\nu}\biggl\lbrace  w({\bf v},{\bf v}_{1};{\bf\Delta})\biggl\lbrack
a_{1}b_{1}(ba'-b'a){\partial^{2}f\over\partial v^{\mu}\partial v^{\nu}}
-(b_{1}a'_{1}-b'_{1}a_{1})(ab'+a'b){\partial f\over\partial v^{\mu}}{\partial f_{1}\over\partial v_{1}^{\nu}}\nonumber\\
+a_{1}b_{1}(ba''-ab'') {\partial f\over\partial v^{\mu}}{\partial f\over\partial v^{\nu}}\biggr\rbrack
+{\partial w\over\partial v^{\mu}}\biggl\lbrack a_{1}b_{1}(ba'-b'a){\partial f\over\partial v^{\nu}}-ab(b_{1}a'_{1}-a_{1}b'_{1}){\partial f_{1}\over\partial v_{1}^{\nu}}\biggr\rbrack\biggr\rbrace.
\end{eqnarray}
Equation (\ref{wd8}) can be written more compactly as
\begin{eqnarray}
\label{wd9}
{df\over dt}={\partial\over\partial v^{\mu}}\int  d^{3}{\bf v}_{1}\ K^{\mu\nu}\biggl\lbrace
a_{1}b_{1}(ba'-b'a){\partial f\over\partial v^{\nu}}-a b (b_{1}a'_{1}-b'_{1}a_{1}){\partial f_{1}\over\partial v_{1}^{\nu}}\biggr\rbrace,
\end{eqnarray}
where we have defined
\begin{equation}
\label{wd10}
K^{\mu\nu}={1\over 2}\int \ d\Omega \ w({\bf v},{\bf v}_{1};{\bf\Delta})\Delta^{\mu}\Delta^{\nu}.
\end{equation}
We now define the functions $g$ and $h$ by
\begin{equation}
\label{wd11}
g(f)=a(f)b(f), \qquad h(f)=b(f)a'(f)-b'(f)a(f).
\end{equation}
Then, Eq. (\ref{wd9}) can be rewritten
\begin{eqnarray}
\label{wd12}
{df\over  dt}={\partial\over\partial v^{\mu}}\int  d^{3}{\bf v}_{1}\ K^{\mu\nu}\biggl\lbrace
g(f_{1})h(f){\partial f\over\partial v^{\nu}}-g(f)h(f_{1}){\partial f_{1}\over\partial v_{1}^{\nu}}\biggr\rbrace,
\end{eqnarray}
This will be called the generalized Landau equation. The tensor
$K^{\mu\nu}$ can be calculated explicitly in the linear trajectory
approximation \cite{balescu}. The result can be expressed as
\begin{equation}
\label{wd13}
K^{\mu\nu}={A\over u}\biggl (\delta^{\mu\nu}-{u^{\mu}u^{\nu}\over u^{2}}\biggr ),
\end{equation}
where ${\bf u}={\bf v}_{1}-{\bf v}$ is the relative velocity and
$A$ is a constant (in the plasma case $A=(e^4/8\pi m^3
\epsilon_0^2)\ln (L_{Debye}/L_{min}$) and in the gravitational
case $A=2\pi G^2 m \ln (L_{max}/L_{min})$).

By developing a kinetic theory of 2D point vortices
\cite{kin,houches}, we have derived a kinetic equation of the form
\begin{equation}
\label{kq1}
{\partial P\over\partial t}+\langle {\bf V}\rangle\nabla P={N\gamma^{2}\over 8}{\partial\over\partial r^{\mu}}\int d^{2}{\bf r}_{1} K^{\mu\nu}\delta({\mb\xi}\cdot {\bf v})\biggl ( P_{1}{\partial P\over\partial r^{\nu}}-P{\partial P_{1}\over\partial r_{1}^{\nu}}\biggr ),
\end{equation}
where
\begin{equation}
\label{kq2}
K^{\mu\nu}={\xi^{2}\delta^{\mu\nu}-\xi^{\mu}\xi^{\nu}\over \xi^{2}},
\end{equation}
and ${\mb\xi}={\bf r}_{1}-{\bf r}$, ${\bf v}=\langle {\bf
V}\rangle({\bf r}_{1},t)-\langle {\bf V}\rangle ({\bf r},t)$. This
equation conserves all the constraints of the point vortex model
and increases the Boltzmann entropy ($H$-theorem). It is
reminiscent of the Landau equation but it {\it differs} from the
Landau equation due to the $\delta$-function which takes into
account the conservation of energy $E={1\over 2}\int \omega\psi
d^{2}{\bf r}$ where $\psi({\bf r},t)$ is the stream-function. In
addition, its physical interpretation and derivation is completely
different from that of the Landau equation. We can heuristically
propose a formal extension of this equation to the more general
form
\begin{equation}
\label{kq3} {\partial P\over\partial t}+\langle {\bf
V}\rangle\nabla P= {N\gamma^{2}\over 8}{\partial\over\partial
r^{\mu}} \int d^{2}{\bf r}_{1} K^{\mu\nu}\delta({\mb\xi}\cdot {\bf
v}) \biggl ( g(P_{1})h(P){\partial P\over\partial
r^{\nu}}-g(P)h(P_{1}) {\partial P_{1}\over\partial
r_{1}^{\nu}}\biggr ).
\end{equation}

\subsection{Generalized entropy}
\label{sec_ge}

We shall say that a kinetic equation possesses a generalized
microcanonical thermodynamical structure if it conserves mass and
energy and increases continuously a functional of the form
\begin{equation}
\label{ge1}
S=-\int C(f)d^{3}{\bf r}d^{3}{\bf v},
\end{equation}
where $C(f)$ is a convex function. By analogy with ordinary
thermodynamics, the functional (\ref{ge1}) will be called a {\it
generalized entropy} even if it does not correspond to a true entropy
in the strict sense (in that case, we speak of a thermodynamical
analogy). The equilibrium distribution function $f_{eq}({\bf r},{\bf
v})$, reached by the kinetic equation for $t\rightarrow +\infty$,
maximizes $S[f]$ at fixed $E$ and $M$. Introducing Lagrange
multipliers and writing the variational principle in the form
\begin{equation}
\label{ge2}
\delta S-\beta\delta E-\alpha\delta M+\beta {\bf \Omega}\delta {\bf L}+\beta {\bf U}\delta{\bf P}=0,
\end{equation}
where we have also accounted for the conservation of angular momentum and impulse, we find that the equilibrium distribution function is given by
\begin{equation}
\label{ge3} C'(f_{eq})=-\beta \biggl ({v^{2}\over 2}+\Phi\biggr
)+\beta {\bf \Omega}({\bf r}{\times} {\bf v})+\beta {\bf U}{\bf
v}-\alpha.
\end{equation}

As we shall see, the generalized Landau equation (\ref{wd12}) possesses
a microcanonical thermodynamical structure. The generalized entropy can be
expressed in terms of the functions $a(f)$ and $b(f)$ as
\begin{equation}
\label{ge4}
S=-\int C(f)d^{3}{\bf r}d^{3}{\bf v},\qquad {\rm
with}\qquad  C'(f)=\ln\biggl\lbrack {a(f)\over b(f)}\biggr\rbrack.
\end{equation}
From Eqs. (\ref{ge4}) and (\ref{wd11}), we obtain the relation
\begin{equation}
\label{ge5} h(f)=g(f)C''(f),
\end{equation}
which leads to the identities
\begin{equation}
\label{ge6}
a(f)=\sqrt{g(f)}e^{{1\over 2}C'(f)}, \qquad b(f)=\sqrt{g(f)}e^{-{1\over 2}C'(f)}.
\end{equation}
If we know $a$ and $b$, we can obtain $g$ and $h$ from
Eq. (\ref{wd11}) and $C$ from Eq. (\ref{ge4}). Alternatively, if we
know $h$ and $g$, we obtain $C''(f)$ from Eq. (\ref{ge5}) and deduce
$a$ and $b$ from Eq. (\ref{ge6}) (up to a multiplicative factor). If
we only specify $C(f)$, we cannot obtain $a$ and $b$ individually but
only the ratio $a/b$. In the next section, we shall consider
simplified forms of the generalized Landau equation where everything
is determined by the specification of the generalized entropy $S[f]$,
or equivalently by  the function $C(f)$.

\subsection{Simplified forms of the generalized Landau equation}
\label{sec_s}

We shall first impose that
\begin{equation}
\label{s1}
g(f)=f, \qquad  h(f)=fC''(f),
\end{equation}
where $C$ is a convex function. In that case, Eq. (\ref{wd12}) takes the
 form
\begin{eqnarray}
\label{s2} {df\over dt}={\partial\over\partial
v^{\mu}}\int  d^{3}{\bf v}_{1}\ K^{\mu\nu} f f_{1}\biggl\lbrace
C''(f){\partial f\over\partial v^{\nu}}-C''(f_{1}){\partial
f_{1}\over\partial v_{1}^{\nu}} \biggr\rbrace.
\end{eqnarray}
This generalized Landau equation was first written in
\cite{gt}. From Eqs. (\ref{ge6}) and (\ref{s1}) we find that $a(f)$ and
$b(f)$ are related to $C(f)$ by
\begin{eqnarray}
\label{s3} a(f)=\sqrt{f}e^{{1\over 2}C'(f)}, \qquad
b(f)=\sqrt{f}e^{-{1\over 2}C'(f)}.
\end{eqnarray}
It is interesting to consider particular cases of the generalized
Landau equation (\ref{s2}). For the Boltzmann entropy $C(f)=f\ln f$, we
recover the ordinary Landau equation
\begin{eqnarray}
\label{s4} {df\over dt}={\partial\over\partial
v^{\mu}}\int  d^{3}{\bf v}_{1}\ K^{\mu\nu}\biggl ( f_{1}{\partial
f\over\partial v^{\nu}}-f{\partial f_{1}\over\partial
v_{1}^{\nu}}\biggr ),
\end{eqnarray}
which is the weak deflexion limit of the ordinary Boltzmann
equation (\ref{bol1}) corresponding to
\begin{eqnarray}
\label{s5}
\qquad a(f)=f, \qquad b(f)=1,
\end{eqnarray}
in Eq. (\ref{bol2}). For the Tsallis entropy $C(f)={1\over
q-1}(f^{q}-f)$, we obtain the $q$-Landau equation
\begin{eqnarray}
\label{s6} {df\over dt}={\partial\over\partial
v^{\mu}}\int  d^{3}{\bf v}_{1}\ K^{\mu\nu}\biggl (f_1 {\partial
f^q\over\partial v^{\nu}}-f {\partial f_{1}^q\over\partial
v_{1}^{\nu}}\biggr ),
\end{eqnarray}
which is associated to the $q$-Boltzmann equation (\ref{bol2})
with
\begin{eqnarray}
\label{s7}
a(f)=\sqrt{f}e^{{1\over 2(q-1)}(q f^{q-1}-1)}, \qquad b(f)=\sqrt{f}e^{-{1\over 2(q-1)}(q f^{q-1}-1)}.
\end{eqnarray}

Instead of Eq. (\ref{s1}), we can impose the relations
\begin{equation}
\label{s8} h(f)=1, \qquad  g(f)=1/C''(f).
\end{equation}
In that case, Eq. (\ref{wd12}) reduces to
\begin{eqnarray}
\label{s9}
{df\over dt}={\partial\over\partial v^{\mu}}\int  d^{3}{\bf v}_{1}\ K^{\mu\nu} \biggl\lbrace
{1\over C''(f_{1})}{\partial f\over\partial v^{\nu}}-{1\over C''(f)}{\partial f_{1}\over\partial v_{1}^{\nu}}\biggr\rbrace.
\end{eqnarray}
This alternative form was also written in \cite{gt}. From Eqs.
(\ref{ge6}) and (\ref{s8}), we find that $a(f)$ and $b(f)$ are
related to $C(f)$ by
\begin{eqnarray}
\label{s10}
a(f)={1\over \sqrt{C''(f)}}e^{{1\over 2}C'(f)}, \qquad b(f)={1\over \sqrt{C''(f)}}e^{-{1\over 2}C'(f)}.
\end{eqnarray}
For the entropy $C(f)=f\ln f+(\eta_{0}-\mu f)\ln (\eta_{0}-\mu f)$, we get
\begin{eqnarray}
\label{s11} {df\over dt}={\partial\over\partial v^{\mu}}\int
d^{3}{\bf v}_{1}\ K^{\mu\nu} \biggl\lbrace f_1 (\eta_{0}-\mu f_1)
{\partial f\over\partial v^{\nu}}-f (\eta_{0}-\mu f){\partial
f_{1}\over\partial v_{1}^{\nu}}\biggr\rbrace.
\end{eqnarray}
For $\mu=1$, the foregoing equation describes the case of fermions
accounting for the Pauli exclusion principle. For $\mu=1$, it
describes the case of bosons. For intermediate values of $\mu$ it
describes a intermediate quantum statistics interpolating between
the Bose and Fermi ones. The generalized Landau equation
(\ref{s11}) is the weak deflexion limit of the generalized
Boltzmann equation (\ref{bol2}) with
\begin{eqnarray}
\label{s12} a(f)=f, \qquad b(f)=\eta_{0}-\mu f.
\end{eqnarray}

Although this extension of kinetic theory is interesting on a formal
point of view, we do {\it not} claim that it is relevant to plasma
physics and stellar dynamics (except in the quantum case).  Indeed, in
neutral plasmas where the interaction is short-ranged due to Debye
shielding the diffusion is normal and the ordinary Landau equation is
rigorously valid. On the other hand, in collisional stellar systems,
the diffusion is only {\it slightly} anomalous due to logarithmic
divergences \cite{lee}. Therefore, the ordinary Landau equation
remains marginally valid when logarithmic divergences are properly
regularized \cite{kandrupR}.  There can be corrections to the ordinary
Landau equation due to memory effects and spatial delocalization
\cite{kandrup} (similar effects arise in the kinetic theory of point
vortices \cite{kin}).  However, this does {\it not} apparently justify
a rigorous notion of generalized thermodynamics (even if deviations to
the Maxwellian distribution may be expected for {\it intermediate}
times). For large times, ordinary thermodynamics (based on the
Boltzmann entropy) is rigorously justified for collisional stellar
systems (and point vortices) in a suitable thermodynamic limit
although these systems are non-extensive and non-additive (see
\cite{grand} for a more complete discussion). Therefore, there does
not seem to be any justification of a generalized thermodynamics for
Hamiltonian systems of point particles in the infinite time limit
(collisional relaxation). At the present time, it is not clear to
which systems the generalized kinetic theory developed previously
could apply (with the exception of quantum particles). This is an open
problem left for future investigations. Our guess is that generalized
kinetic equations can serve as {\it effective equations} in the case
of complicated systems. In the following section, we show that a
notion of generalized thermodynamics also emerges in the context of
the violent relaxation of collisionless stellar systems (and other
Hamiltonian systems with long-range interactions) for a completely
different reason. Generalized kinetic equations can find physical
applications in that context.

\section{Violent relaxation of collisionless stellar systems}
\label{sec_vr}

\subsection{The Vlasov-Poisson system}
\label{sec_vp}

For most stellar systems, including the important class of
elliptical galaxies, the encounters between stars are completely
negligible \cite{bt,dubrovnik} and the galaxy dynamics is described by the
self-consistent Vlasov-Poisson system
\begin{equation}
\label{Vlasov}
{\partial f\over\partial t}+{\bf v}{\partial f\over\partial {\bf r}}+{\bf F}{\partial f\over\partial {\bf v}}=0,
\end{equation}
\begin{equation}
\label{Poisson}
\Delta\Phi=4\pi G\int f d^{3}{\bf v}.
\end{equation}
Here, $f({\bf r},{\bf v},t)$ denotes the distribution function
(defined such that $f d^{3}{\bf r}d^{3}{\bf v}$ gives the total
mass of stars with position ${\bf r}$ and velocity ${\bf v}$ at
time $t$), ${\bf F}({\bf r},t)=-\nabla\Phi$ is the gravitational
force (by unit of mass) experienced by a star and $\Phi({\bf
r},t)$ is the gravitational potential related to the star density
$\rho({\bf r},t)=\int f d^{3}{\bf v}$ by the Newton-Poisson
equation (\ref{Poisson}). The Vlasov equation (\ref{Vlasov})
simply states that, in the absence of encounters, the distribution
function $f$ is conserved by the flow in phase space. This can be
written $df/dt=0$ where $d/dt={\partial /\partial t}+{\bf
U}_{6}\nabla_{6}$ is the material derivative and ${\bf
U}_{6}=({\bf v},{\bf F})$ is a generalized velocity field in the
$6$-dimensional phase space $({\bf r},{\bf v})$ (by definition,
$\nabla_{6}=(\partial/\partial {\bf r},\partial/\partial {\bf v})$
is the generalized nabla operator). Since the flow is
incompressible, i.e. $\nabla_{6}{\bf U}_{6}=0$, the hypervolume of
a ``fluid'' particle is conserved. Since, in addition, a fluid
particle conserves the distribution function, this implies that
the total mass (or hypervolume) of all phase elements with phase
density between $f$ and $f+\delta f$ is conserved. This is
equivalent to the conservation of the Casimir integrals $
I_{h}=\int h(f) d^{3}{\bf  r} d^{3}{\bf  v}$ for any continuous
function $h(f)$. This is also equivalent to the conservation of
the moments
\begin{equation}
\label{E15} M_n=\int f^{n} d^{3}{\bf r}d^{3}{\bf v},
\end{equation}
which include in particular the total mass $M=\int f d^{3}{\bf r}
d^{3}{\bf v}$. It is also straightforward to check that the
Vlasov-Poisson system conserves the total energy $E$, the angular
momentum ${\bf L}$ and the impulse ${\bf P}$.

\subsection{The metaequilibrium state}
\label{sec_meta}

The Vlasov-Poisson system develops very complex filaments as a result
of a mixing process in phase space. If we introduce a coarse-graining
procedure, the coarse-grained distribution function $\overline{f}({\bf
r},{\bf v},t)$ will reach a metaequilibrium state $\overline{f}({\bf
r},{\bf v})$ on a very short timescale, of the order of the dynamical
time. This process is known as ``phase mixing'' and ``violent relaxation''
\cite{bt}. Lynden-Bell \cite{lb} has tried to describe this
metaequilibrium state in terms of statistical mechanics. If $\rho({\bf
r},{\bf v},\eta)$ denotes the density probability of finding the value
$f=\eta$ of distribution function in $({\bf r}, {\bf v})$, then the
mixing entropy is given by
\begin{equation}
\label{meta1} S[\rho]=-\int \rho \ln\rho \ d^{3}{\bf r} d^{3}{\bf v} d\eta.
\end{equation}
It can be obtained by a standard combinatorial analysis \cite{lb} or by
using the concept of Young measures and large deviations
\cite{mr}. Assuming ergodicity (which may not be realized in practice,
see below) the statistical equilibrium state is obtained by maximizing
$S[\rho]$ while conserving mass $M$, energy $E$ and all the Casimirs
(or moments $M_n$). The optimal equilibrium state can be written
\cite{lb,grand}
\begin{equation}
\label{meta2} \rho({\bf r},{\bf v},\eta) ={1\over
Z}\chi(\eta)e^{-(\beta\epsilon+\alpha)\eta},
\end{equation}
where $\epsilon={v^2\over 2}+\Phi$ is the energy of a star by unit
of mass, $\chi(\eta)\equiv {\rm
exp}(-\sum_{n>1}\alpha_{n}\eta^{n})$ accounts for the conservation
of the fragile moments $M_{n>1}=\int \rho \eta^{n}d\eta d^{3}{\bf
r}d^{3}{\bf v}$ and $\alpha$, $\beta$ are Lagrange multipliers for
$M$ and $E$ (robust integrals). The partition function $Z$ is
determined by the local normalization $\int \rho d\eta=1$ and the
equilibrium coarse-grained field is given by $\overline{f}=\int
\rho \eta d\eta$. This can be written \cite{grand}
\begin{equation}
\label{meta2b}
\overline{f}=-{1\over\beta}{\partial\ln Z\over\partial \epsilon}=F(\beta\epsilon+\alpha)=\overline{f}(\epsilon).
\end{equation}
We emphasize that Eq. (\ref{meta2b}) is obtained after two
successive coarse-grainings. Intrinsically, a galaxy is a
collection of $N$ point-like stars and the exact distribution
function $f_{exact}=\sum_{i=1}^{N}m_{i}\delta ({\bf r}-{\bf
r}_{i})\delta({\bf v}-{\bf v}_{i})$ is a sum of
$\delta$-functions. As is customary in statistical mechanics, we
consider a smooth distribution function $f({\bf r},{\bf v},t)$
which is the statistical average of $f_{exact}$, i.e. $f=\langle
f_{exact}\rangle$. In the collisionless regime valid for times
$t\ll t_{coll}\sim {N\over\ln N}t_{dyn}$ ($t_{coll}$ is the
timescale of collisional relaxation and $t_{dyn}$ the dynamical
time), this function $f({\bf r},{\bf v},t)$ satisfies the Vlasov
equation (\ref{Vlasov}). However, the Vlasov-Poisson system
develops a mixing process and it is relevant to introduce {\it
another} coarse-graining (at a much larger scale) and define
$\overline{f}({\bf r},{\bf v},t)$ as the local average of $f({\bf
r},{\bf v},t)$ on a phase space cell of volume
$\epsilon_{r}^{3}\epsilon_{v}^{3}$. The quantity appearing in Eq.
(\ref{meta2b}) is the most probable form of $\overline{f}$ at
statistical equilibrium (in the sense of Lynden-Bell), assuming
ergodicity (i.e. efficient mixing).

From Eq. (\ref{meta2}), it is easy to show
 (see the equivalent proof in \cite{sw}) that
\begin{equation}
\label{meta3} \overline{f}'(\epsilon)=-\beta f_{2},\qquad  f_{2}\equiv \int \rho (\eta-\overline{f})^{2}d\eta>0,
\end{equation}
where $f_{2}$ is the centered local variance of the distribution
$\rho({\bf r},{\bf v},\eta)$. Therefore,
$\overline{f}=\overline{f}(\epsilon)$ is a decreasing function of
the stellar energy (assuming $\beta>0)$. Since
$\overline{f}(\epsilon)$ is monotonic, the coarse-grained
distribution  (\ref{meta2b})  extremizes a ``generalized entropy''
\cite{grand,gt}
\begin{equation}
\label{meta4} S[f]=-\int C(f) d^{3}{\bf r}d^{3}{\bf v},
\end{equation}
at fixed mass $M$ and energy $E$, where $C(f)$ is a convex function,
i.e. $C''>0$. Indeed, introducing Lagrange multipliers and writing the
variational principle in the form
\begin{equation}
\label{meta5} \delta S-\beta \delta E-\alpha\delta M=0,
\end{equation}
we find that
\begin{equation}
\label{meta6} C'(\overline{f})=-\beta\epsilon-\alpha.
\end{equation}
Since $C'$ is a monotonically increasing function of $f$, we can
inverse this relation to obtain
\begin{equation}
\label{meta7} \overline{f}=F(\beta\epsilon+\alpha)=\overline{f}(\epsilon),
\end{equation}
where $F(x)=(C')^{-1}(-x)$. Equation (\ref{meta7}) can be compared to
Eq. (\ref{meta2b}). From the identity
\begin{equation}
\label{meta7b} \overline{f}'(\epsilon)=-\beta/C''(\overline{f}),
\end{equation}
resulting from Eq. (\ref{meta6}), $\overline{f}(\epsilon)$ is a
monotonically decreasing function of energy if $\beta>0$. Therefore,
for any Gibbs state of the form (\ref{meta2}), there exists a
generalized entropy of the form (\ref{meta4}) that $\overline{f}$
extremizes (at fixed $E$, $M$). It can be shown furthermore that
$\overline{f}$ {\it maximizes} this functional. We note that $C({f})$,
hence the generalized entropy (\ref{meta4}), is a {\it non-universal}
function which depends on the initial conditions. In general, it is
{\it not} the ordinary Boltzmann entropy $S_B[\overline{f}]=-\int
\overline{f}\ln \overline{f} d^{3}{\bf r}d^{3}{\bf v}$ due to
fine-grained constraints (Casimirs) that modify the form of entropy
that we would naively expect. We emphasize that maximizing the multi-levels
Boltzmann entropy $S[\rho]$ at fixed mass $M$, energy $E$ and with an
infinite number of fine-grained constraints $M_{n}$ (Casimirs) gives
the same result for the coarse-grained field $\overline{f}$ as
maximizing a certain generalized entropy $S[\overline{f}]$
(non-universal) while conserving only mass $M$ and energy $E$ (robust
constraints). The existence of ``hidden constraints'' (here the
Casimir invariants that are not accessible at the coarse-grained
scale) is the physical reason for the occurence of ``generalized
entropy'' functionals in a problem. {\it We can either work with the
Boltzmann entropy and take into account all the constraints imposed by
the dynamics, or keep only the constraints that are the most directly
accessible to the observations and change the form of entropy}. This
clearly leads to an indetermination which appears in the parameter $q$
of Tsallis or more generally in our function $C(f)$.  We expect this
idea of ``hidden constraints'' to be very general and of fundamental
importance.

The above statistical approach rests on the assumption that the
evolution is ergodic. In reality, this is not the case. It has
been understood since the beginning \cite{lb} that violent
relaxation is {\it incomplete} so that the Boltzmann entropy
(\ref{meta1}) is {\it not} maximized in the whole available phase
space (this is independant on the fact that it has {no} maximum!).
However, {\it the metaequilibrium state reached by the system as a
result of incomplete violent relaxation is always a nonlinearly
dynamically stable solution of the Vlasov-Poisson system (on the
coarse-grained scale)}. If $\overline{f}=\overline{f}(\epsilon)$,
which is a particular case of the Jeans theorem \cite{bt}, it
maximizes a functional of the form (\ref{meta4}) at fixed mass and
energy. In this dynamical context, $S[f]$ is called a $H$-function
\cite{tremaine}. This functional depends on the initial conditions
(for the same reasons as before) and also on the strength of
mixing (if the mixing is complete, $C(f)$ can be predicted by the
statistical theory of Lynden-Bell). Boltzmann and Tsallis
functionals are particular H-functions corresponding to isothermal and
polytropic distribution functions. They are not good models of
incomplete relaxation for elliptical galaxies \cite{grand}. A
better model is a {\it composite model} that is isothermal in the
core and polytropic in the halo. Since the variational principle
determining the nonlinear dynamical stability of a collisionless
stellar system (maximization of a H-function at fixed mass and
energy) is {\it similar} to the usual thermodynamical variational
principle (maximization of the Boltzmann entropy at fixed mass and
energy) we can use a {\it thermodynamical analogy} to analyze the
dynamical stability of collisionless stellar systems
\cite{grand,gt}. In this analogy, the H-function can be called a
``generalized entropy''.  We believe that this {\it
thermodynamical analogy}  is the correct
interpretation of the notion of ``generalized thermodynamics''
introduced by Tsallis in the context of stellar systems (and 2D
turbulence). We emphasize that the maximization of a $H$-function
(e.g., Tsallis entropy) at fixed mass and energy is a condition of
nonlinear dynamical stability, not a condition of thermodynamical
stability.

\subsection{Heuristic approach of violent relaxation}
\label{sec_heuristic}

Violent relaxation is a very complicated concept because of the
presence of fine-grained constraints (Casimirs). The Casimirs differ
from robust constraints such as mass and energy because they are
altered by the coarse-graining procedure since $\overline{f^n}\neq
\overline{f}^{n}$ for $n\neq 1$. Therefore, they can be determined
only from the initial conditions (which are not mixed) at $t=0$, say.
Unfortunately, in practice, we do not know the initial conditions
(e.g., the initial state that gave rise to an elliptical galaxy) so
that we do not know the Casimirs. Only mass and energy (robust
constraints) can be determined at all times $t\ge 0$ since
$\overline{M}=M$ and $\overline{E}\simeq E$ \cite{grand}.  We thus
have to deal with a limited amount of information on the
system. Therefore, {\it we cannot make predictions because we do not
know all the constraints on the system}. If we want to make some
predictions, we have two possibilities: (i) the first possibility is
to ``guess'' the initial conditions (consistent with the information
that we have on the system) and determine the corresponding
equilibrium state. We can then study how the equilibrium state depends
on the initial conditions (for given $E$ and $M$). (ii) the second
possibility is to ``guess'' the generalized entropy that is maximized
by the system at equilibrium. We can then study how the equilibrium
state depends on the form of $C(f)$ and whether generalized entropies
presenting the same properties can be regrouped in ``classes of
equivalence''
\cite{gt}.

Each approach has its own advantages and drawbacks. The first approach
is more complicated because, for realistic initial conditions, we have
to account for an infinity of constraints (Casimirs) in addition to
mass and energy. This clearly leads to practical difficulties. In
addition, it is not clear whether all these constraints are physically
relevant because they can be altered by non-ideal effects as discussed
in \cite{grand,gt}. Finally, this description assumes that the
mixing is complete which is not the case in practice.  We shall
therefore prefer the second approach where we have only two robust
constraints $M$ and $E$. The other constraints are taken into account
implicitly in the form of the generalized entropy that we
consider. This is a much more convenient approach. In addition, this
approach is directly connected to the problem of dynamical stability
that we discussed previously. It has therefore a lot of attractive
advantages.

\subsection{The quasilinear theory}
\label{sec_qt}

Basically, a collisionless stellar system is described in a
self-consistent mean field approximation by the Vlasov-Poisson system
(\ref{Vlasov})(\ref{Poisson}). In principle, these coupled equations
determine completely the evolution of the distribution function
$f({\bf r},{\bf v},t)$. However, as discussed in Sec. \ref{sec_meta},
we are not interested in practice by the finely striated structure of
the flow in phase space but only by its macroscopic,
i.e. smoothed-out, structure. Indeed, the observations and the
numerical simulations are always realized with a finite
resolution. Moreover, the ``coarse-grained'' distribution function
$\overline{f}({\bf r},{\bf v},t)$ is likely to converge towards an
equilibrium state contrary to the exact distribution $f$ which
develops smaller and smaller scales.

If we decompose the distribution function and the gravitational
potential in a mean and fluctuating part ($f=\overline{f}+\tilde f$,
$\Phi=\overline{\Phi}+\tilde\Phi$) and take the local average of the
Vlasov equation (\ref{Vlasov}), we readily obtain an equation of the
form
\begin{equation}
\label{E15b}
{\partial\overline{f}\over\partial t}+{\bf v}{\partial \overline{f}\over\partial {\bf r}}+
\overline{\bf F}{\partial \overline{f}\over\partial {\bf v}}=
-{\partial {\bf J}_{f}\over\partial {\bf v}},
\end{equation}
for the ``coarse-grained'' distribution function with a diffusion
current ${\bf J}_{f}=\overline{\tilde f\tilde {\bf F}}$ related to
the correlations of the ``fine-grained'' fluctuations.  Any
systematic calculation of the diffusion current starting from the
Vlasov equation (\ref{Vlasov}) must necessarily introduce an
evolution equation for $\tilde f$. This equation is simply
obtained by subtracting Eq. (\ref{E15b}) from Eq.
(\ref{Vlasov}). This yields
\begin{equation}
\label{E29}
{\partial \tilde{f}\over\partial t}+{\bf v}{\partial \tilde{f}\over \partial {\bf r}}+\overline{\bf F}{\partial \tilde{f}\over\partial {\bf v}}=-\tilde{\bf F}{\partial \overline{f}\over\partial {\bf v}}-\tilde {\bf F}{\partial \tilde f\over\partial {\bf v}}+\overline{\tilde {\bf F}{\partial \tilde f\over\partial {\bf v}}}.
\end{equation}

To go further, we need to implement some approximations. In the
sequel, we shall develop a quasilinear theory (see also
\cite{kp,sl,dubrovnik}). This will provide a precise theoretical
framework to analyze the process of ``collisionless relaxation''
in stellar systems.  The essence of the quasilinear theory is to
assume that the fluctuations are weak and neglect the nonlinear
terms in Eq. (\ref{E29}) altogether. In that case, Eqs.
(\ref{E15b}) and (\ref{E29}) reduce to the coupled system
\begin{equation}
\label{E30}
{\partial \overline{f}\over\partial t}+L\overline{f}=-{\partial \over \partial {\bf v}}\overline{\tilde {\bf F}\tilde f},
\end{equation}
\begin{equation}
\label{E31}
{\partial \tilde{f}\over\partial t}+L\tilde{f}=-\tilde {\bf F}{\partial\overline{f}\over\partial {\bf v}},
\end{equation}
where $L={\bf v}{\partial\over\partial {\bf r}}+\overline{\bf
F}{\partial\over\partial {\bf v}}$ is the advection operator in phase
space.  Physically, these equations describe the coupling between a
subdynamics (here the small scale fluctuations $\tilde f$) and a
macrodynamics (described by the coarse-grained distribution function
$\overline{f}$). Due to the strong simplifications implied by the
neglect of nonlinear terms in Eq. (\ref{E29}), the quasilinear theory
only describes the late quiescent stages of the violent relaxation
process, when the fluctuations have weaken (gentle
relaxation). Although this is essentially an asymptotic theory, it is
of importance to develop this theory in detail since it provides an
explicit expression for the effective ``collision operator'' which
appears on a coarse-grained scale.

Introducing the Greenian
\begin{equation}
\label{E32}
{\cal G}(t_{2},t_{1})\equiv \exp \Biggl \lbrace -\int_{t_{1}}^{t_{2}}dt L(t)\Biggr \rbrace,
\end{equation}
we can immediately write down a formal solution of Eq. (\ref{E31}), namely
\begin{equation}
\label{E33}
\tilde f ({\bf r},{\bf v},t)={\cal G}(t,0)\tilde f ({\bf r},{\bf v},0)-\int_{0}^{t}ds {\cal G}(t,t-s)\tilde {\bf F}({\bf r},t-s){\partial\overline{f}\over\partial {\bf v}}({\bf r},{\bf v},t-s).
\end{equation}
Although very compact, this formal expression is in fact extremely
complicated. Indeed, all the difficulty is encapsulated in the
Greenian ${\cal G}(t,t-s)$ which supposes that we can solve the smoothed-out
Lagrangian flow
\begin{equation}
\label{E34}
{d{\bf r}\over dt}={\bf v},\qquad {d{\bf v}\over dt}=\overline{\bf F},
\end{equation}
between $t$ and $t-s$. In practice, this is impossible and we will
have to make some approximations.

The objective now is to substitute the formal result (\ref{E33})
back into Eq. (\ref{E30}) and make a closure approximation in
order to obtain a self-consistant equation for $\overline{f}({\bf
r},{\bf v},t)$.  If the fluctuating force $\tilde{\bf F}$ were
external to the system, we would simply obtain a diffusion
equation
\begin{equation}
\label{E35}
{\partial \overline{f}\over\partial t}+L\overline{f}= {\partial\over\partial v^{\mu}}\biggl (D^{\mu\nu}{\partial\overline{f}\over\partial v^{\nu}}\biggr ),
\end{equation}
with a diffusion coefficient given by a Kubo formula
\begin{equation}
\label{E36}
D^{\mu\nu}=\int_{0}^{t} ds \overline{\tilde F^{\mu}({\bf r},t)\tilde F^{\nu}({\bf r},t-s)}.
\end{equation}
However, in the case of the Vlasov-Poisson system, the
gravitational force is produced by the distribution of matter
itself and this coupling will give rise to a friction term in
addition to the pure diffusion. Indeed, we have
\begin{equation}
\label{E37}
\tilde {\bf F}({\bf r},t)=\int {\bf F}({\bf r}'\rightarrow {\bf r})\tilde f ({\bf r}',{\bf v'},t)d^{3}{\bf r}'d^{3}{\bf v}',
\end{equation}
where
\begin{equation}
\label{E38}
{\bf F}({\bf r}'\rightarrow {\bf r})=G {{\bf r}'-{\bf r}\over |{\bf r}'-{\bf r}|^{3}},
\end{equation}
represents the force (by unit of mass) created by a star in ${\bf
r'}$ on a star in ${\bf r}$ (Newton's law). Therefore, considering
Eqs. (\ref{E33}) and (\ref{E37}), we see that the
fluctuations of the distribution function $\tilde f ({\bf r},{\bf
v},t)$ are given by an iterative process: $\tilde f (t)$ depends
on $\tilde {\bf F}(t-s)$ which itself depends on $\tilde f (t-s)$
etc... We shall solve this problem perturbatively in an expansion
in powers of the gravitational constant $G$. This is the
equivalent of the ``weak coupling approximation'' in plasma
physics.

According to Eq. (\ref{E37}), we have
\begin{eqnarray}
\label{E39a}
\overline{\tilde f({\bf r},{\bf v},t)\tilde F^{\mu}({\bf r},t)}=\int d^{3}{\bf r}'d^{3}{\bf v}' F^{\mu}({\bf r}'\rightarrow {\bf r})\overline{\tilde f({\bf r}',{\bf v}',t)\tilde f({\bf r},{\bf v},t)}.
\end{eqnarray}
On the other hand, according to Eqs. (\ref{E33}) and (\ref{E37}), we have
\begin{eqnarray}
\label{E39b}
\tilde f({\bf r},{\bf v},t)={\cal G}(t,0)\tilde f({\bf r},{\bf v},0)-\int_{0}^{t}ds\int d^{3}{\bf r}'' d^{3}{\bf v}''{\cal G}(t,t-s)\nonumber\\
{\times} F^{\nu}({\bf r}''\rightarrow {\bf r})\tilde f({\bf r}'',{\bf v}'',t-s){\partial\overline{f}\over \partial v^{\nu}}({\bf r},{\bf v},t-s),
\end{eqnarray}
and a similar expression for $\tilde f({\bf r}',{\bf v}',t)$. Substituting the foregoing expansion in Eq. (\ref{E39a}), we find that the term of order $G^{2}$  is
\begin{eqnarray}
\label{E39c}
\overline{\tilde f({\bf r},{\bf v},t)\tilde F^{\mu}({\bf r},t)}=-\int_{0}^{t}ds\int d^{3}{\bf r}'' d^{3}{\bf v}''\int d^{3}{\bf r}' d^{3}{\bf v}' F^{\mu}({\bf r}'\rightarrow {\bf r})\nonumber\\
{\times}  \biggl\lbrace {\cal G}'(t,0){\cal G}(t,t-s) F^{\nu}({\bf r}''\rightarrow {\bf r})\overline{\tilde f({\bf r}',{\bf v}',0)\tilde f({\bf r}'',{\bf v}'',t-s)}{\partial\overline{f}\over \partial v^{\nu}}({\bf r},{\bf v},t-s)\nonumber\\
+ {\cal G}(t,0){\cal G}'(t,t-s) F^{\nu}({\bf r}''\rightarrow {\bf r}')\overline{\tilde f({\bf r},{\bf v},0)\tilde f({\bf r}'',{\bf v}'',t-s)}{\partial\overline{f}\over \partial v^{\nu}}({\bf r}',{\bf v}',t-s)\biggr\rbrace.
\end{eqnarray}
Using
\begin{eqnarray}
\label{E39d}
{\cal G}'(t,0)={\cal G}'(t,t-s){\cal G}'(t-s,0),
\end{eqnarray}
and
\begin{eqnarray}
\label{E39e}
\tilde f({\bf r}',{\bf v}',t-s)={\cal G}'(t-s,0)\tilde f({\bf r}',{\bf v}',0)+O(G),
\end{eqnarray}
we obtain after
some rearrangements
\begin{eqnarray}
\label{E39}
{\partial \overline{f}\over\partial t}+L\overline{f}={\partial \over \partial v^{\mu}}\int_{0}^{t} ds\int d^{3}{\bf r'}d^{3}{\bf v'}d^{3}{\bf r''}d^{3}{\bf v''} F^{\mu}({\bf r}'\rightarrow {\bf r}){\cal G}'(t,t-s){\cal G}(t,t-s)\nonumber\\
{\times}\biggl\lbrace F^{\nu}({\bf r}''\rightarrow {\bf r})\overline{ \tilde f({\bf r}',{\bf v}',t-s)\tilde f({\bf r}'',{\bf v}'',t-s)} {\partial \overline{f}\over \partial v^{\nu}}({\bf r},{\bf v},t-s)\nonumber\\
+F^{\nu}({\bf r}''\rightarrow {\bf r}')\overline{\tilde f({\bf r},{\bf v},t-s)\tilde f({\bf r}'',{\bf v}'',t-s)} {\partial \overline{f}\over \partial v'^{\nu}}({\bf r}',{\bf v}',t-s) \biggr \rbrace.\nonumber\\
\end{eqnarray}
In this expression, the Greenian ${\cal G}$ refers to the fluid particle
${\bf r}(t),{\bf v}(t)$ and the Greenian ${\cal G}'$ to the fluid particle
${\bf r}'(t),{\bf v}'(t)$. To close the system, it remains for one to
evaluate the correlation function $\overline{\tilde f({\bf r},{\bf
v},t)\tilde f({\bf r'},{\bf v}',t)}$. We shall assume that the mixing
in phase space is sufficiently efficient so that the scale of the
kinematic correlations is small with respect to the coarse-graining
mesh size. In that case,
\begin{equation}
\label{E40} \overline{\tilde f({\bf r},{\bf v},t)\tilde f({\bf
r'},{\bf v'},t)}=\epsilon_{r}^{3}\epsilon_{v}^{3} \delta({\bf
r}-{\bf r}') \delta ({\bf v}-{\bf v}') f_2({\bf r},{\bf v},t),
\end{equation}
where $\epsilon_{r}$ and $\epsilon_{v}$ are the resolution scales
in position and velocity respectively and
\begin{equation}
\label{E41} f_2\equiv  \overline{\tilde
f^{2}}=\overline{(f-\overline{f})^{2}}=\overline{f^{2}}-\overline{f}^{2}.
\end{equation}
is the local variance of the fine-grained fluctuations.

\subsection{The closure approximation}
\label{sec_c}

We are now led to a closure problem. Indeed, in order to obtain a
self-consistent kinetic equation for $\overline{f}$, we need to
determine the variance $f_{2}$. If the initial condition in phase
space consists of patches of uniform distribution function
$f=\eta_{0}$ surrounded by vacuum $f=0$ (two-levels
approximation), then $\overline{f^{2}}=\overline{\eta_{0}{\times} f}=
\eta_{0}\overline{f}$ and, therefore, $f_{2}= \overline{f}
(\eta_{0}-\overline{f})$. This leads to a generalized Landau
equation of the form (\ref{s11}) as for fermions
\cite{kp,sl,dubrovnik}. For more complicated initial conditions
(multi-levels case), the strategy would be to write down a kinetic
equation for $\rho({\bf r},{\bf v},\eta,t)$, the density
probability of finding the value $f=\eta$ in $({\bf r},{\bf v})$
at time $t$. This extension can be performed along the lines
sketched in \cite{sl} and the resulting equation for $\rho({\bf
r},{\bf v},\eta,t)$ can be closed. However, this approach leads to
a system of $N$ coupled equations (one for each level $\eta$)
which is not convenient to solve when $N\gg 1$. We could
alternatively obtain a hierarchy of equations for the moments
$\overline{f}$, $\overline{f^{2}}$,...,$\overline{f^{n}}$,... but
we would then encounter a closure problem. The equations obtained
by this method are complicated because they take into account the
conservation of all the Casimirs.

In this paper, we propose a closure approximation that leads to a
simpler kinetic equation. While not being exact, this equation
preserves the robust features of the process of violent relaxation and
is amenable to an easier numerical implementation. Its main interest
is to go beyond the two-levels approximation while leaving the problem
tractable. The idea is to observe that Eqs. (\ref{meta3}) and (\ref{meta7b})
lead to the important relation
\begin{equation}
\label{c1} f_{2}={1\over C''(\overline{f})}.
\end{equation}
This relation is valid at equilibrium but we propose to use it as a
closure approximation in Eq. (\ref{E40}). This is expected to be a
reasonable approximation if we are close to equilibrium, which is in
fact dictated by the quasi-linear approximation. Of course, this
procedure assumes that we know the function $C(f)$ in advance. This is
the case if we have already determined the equilibrium state and we
want to describe the dynamics close to equilibrium. This is also the
case in the heuristic approach of violent relaxation discussed in
Sec. \ref{sec_heuristic} where we have to ``guess'' the form of $C(f)$
that is relevant to our system (or try different functionals).

If we substitute Eqs. (\ref{E41})-(\ref{c1}) in Eq. (\ref{E39}) and
carry out the integrations on ${\bf r}''$ and ${\bf v}''$, we obtain
\begin{eqnarray}
\label{E43}
{\partial \overline{f}\over\partial t}+L\overline{f}=\epsilon_{r}^{3}\epsilon_{v}^{3}{\partial \over \partial v^{\mu}}\int_{0}^{t} ds\int d^{3}{\bf r'}d^{3}{\bf v'} F^{\mu}({\bf r}'\rightarrow {\bf r})_{t} F^{\nu}({\bf r}'\rightarrow {\bf r})_{t-s}\nonumber\\
{\times}\biggl\lbrace {1\over C''(\overline{f}')} {\partial \overline{f}\over \partial v^{\nu}}
-{1\over C''(\overline{f}')} {\partial \overline{f}'\over \partial v'^{\nu}} \biggr \rbrace_{t-s}.
\end{eqnarray}
We have written $\overline{f}'_{t-s}\equiv \overline{f}({\bf
r}'(t-s),{\bf v}'(t-s),t-s)$, $\overline{f}_{t-s}\equiv
\overline{f}({\bf r}(t-s),{\bf v}(t-s),t-s)$, $F^{\mu}({\bf
r}'\rightarrow {\bf r})_{t}\equiv F^{\mu}({\bf r}'(t)\rightarrow
{\bf r}(t))$ and $F^{\nu}({\bf r}'\rightarrow {\bf r})_{t-s}\equiv
F^{\nu}({\bf r}'(t-s)\rightarrow {\bf r}(t-s))$ where ${\bf
r}(t-s)$ and ${\bf v}(t-s)$ are the position and velocity at time
$t-s$ of the stellar fluid particle located in ${\bf r}={\bf
r}(t)$, ${\bf v}={\bf v}(t)$ at time $t$. They are determined by
the characteristics (\ref{E34}) of the smoothed-out Lagrangian
flow.

Equation (\ref{E43}) is a non Markovian integrodifferential equation:
the value of $\overline{f}$ in ${\bf r}$, ${\bf v}$ at time $t$
depends on the value of the { whole} field $\overline{f}({\bf r}',{\bf
v}',t-s)$ at { earlier times}. If the decorrelation time $\tau$ is
short, we can make a {\it Markov approximation} and replace the
bracket at time $t-s$ by its value taken at time $t$. Noting
furthermore that the integral is dominated by the contribution of
field stars close to the star under consideration (i.e. when ${\bf
r}'\rightarrow {\bf r}$), we shall make a {\it local approximation}
and replace $C''(\overline{f}')$ and $ {\partial
\overline{f}'\over \partial v'^{\nu}}$ by their values taken at ${\bf
r}$. In that case, the foregoing equation simplifies in
\begin{eqnarray}
\label{E44}
{\partial \overline{f}\over\partial t}+L\overline{f}=\epsilon_{r}^{3}\epsilon_{v}^{3}{\partial \over \partial v^{\mu}}\int_{0}^{t} ds\int d^{3}{\bf r'}d^{3}{\bf v'} F^{\mu}({\bf r}'\rightarrow {\bf r})_{t} F^{\nu}({\bf r}'\rightarrow {\bf r})_{t-s}\nonumber\\
{\times}\biggl\lbrace  {1\over C''(\overline{f}')}{\partial \overline{f}\over \partial v^{\nu}}
-  {1\over C''(\overline{f})} {\partial \overline{f}'\over \partial v'^{\nu}} \biggr \rbrace_{t},
\end{eqnarray}
where, now, $\overline{f}'=\overline{f}({\bf r},{\bf v}',t)$. The
explicit reference to the past evolution of the system is only
retained in the memory function $$\int_{0}^{t} ds \int d^{2}{\bf r}'
F^{\mu}({\bf r}'\rightarrow {\bf r})_{t} F^{\nu}({\bf r}'\rightarrow
{\bf r})_{t-s}.$$ This function can be calculated explicitly if we
assume that, between $t-s$ and $t$, the stars follow linear
trajectories, so that ${\bf v}(t-s)={\bf v}$ and ${\bf r}(t-s)={\bf
r}-{\bf v}s$ \cite{balescu}. This leads to the
generalized Landau equation
\begin{eqnarray}
\label{E45}
{\partial \overline{f}\over\partial t}+L\overline{f}={\partial\over\partial v^{\mu}}\int d^{3}{\bf v}' K^{\mu\nu}\biggl\lbrace {1\over C''(\overline{f}')}{\partial\overline{f}\over\partial v^{\nu}}- {1\over C''(\overline{f}')}{\partial\overline{f}'\over\partial v'^{\nu}}\biggr\rbrace,
\end{eqnarray}
where $K^{\mu\nu}$ is the tensor
\begin{eqnarray}
\label{E46}
K^{\mu\nu}=2\pi G^{2}\epsilon_{r}^{3}\epsilon_{v}^{3}\ln\Lambda {1\over u}\biggl (\delta^{\mu\nu}-{u^{\mu}u^{\nu}\over u^{2}}\biggr ),
\end{eqnarray}
and ${\bf u}={\bf v}'-{\bf v}$, $\ln\Lambda=\ln (R/\epsilon_{r})$.
This equation applies to inhomogeneous systems but, as a result of the
local approximation, the effect of inhomogeneity is only retained in
the advective term.

By using a different approach based on a Maximum Entropy Production
Principle (MEPP), Chavanis, Sommeria \& Robert \cite{csr} have proposed a
relaxation equation for $\rho({\bf r},{\bf v},\eta,t)$ of the form
\begin{eqnarray}
\label{w1}
{\partial\rho\over\partial t}+L\rho={\partial\over\partial {\bf v}}\biggl\lbrace D\biggl\lbrack {\partial\rho\over\partial {\bf v}}+\beta (\eta-\overline{f})\rho {\bf v}\biggr\rbrack\biggr\rbrace.
\end{eqnarray}
From this equation, we can deduce a hierarchy of equations for the moments $\overline{f^{n}}=\int \rho \eta^{n} d^{3}{\bf r}d^{3}{\bf v}d\eta$. The equation for the first moment $\overline{f}$ is
\begin{eqnarray}
\label{w2}
{\partial\overline{f}\over\partial t}+L\overline{f}={\partial\over\partial {\bf v}}\biggl\lbrace D\biggl\lbrack {\partial\overline{f}\over\partial {\bf v}}+\beta f_{2} {\bf v}\biggr\rbrack\biggr\rbrace.
\end{eqnarray}
If we close the hierarchy of equations with Eq. (\ref{c1}), we obtain a self-consistent equation of the form
\begin{eqnarray}
\label{w3}
{\partial\overline{f}\over\partial t}+L\overline{f}={\partial\over\partial {\bf v}}\biggl\lbrace D\biggl\lbrack {\partial\overline{f}\over\partial {\bf v}}+{\beta \over C''(\overline{f})}  {\bf v}\biggr\rbrack\biggr\rbrace.
\end{eqnarray}
Note that this equation can also be derived from Eq. (\ref{E45}) by
replacing $\overline{f}'$ by its equilibrium value. Then, the
diffusion coefficient $D$ can be explicitly evaluated (see
\cite{mnras} in the two-levels approximation). Furthermore,
Chavanis {\it et al.} \cite{csr} proposed to let $\beta$ depend on
time, i.e. $\beta=\beta(t)$, so as to conserve energy. This
heuristic ``microcanonical description'' is more adapted to the
context of the violent relaxation than a ``canonical description'' with fixed $\beta$.

By developing a quasi-linear theory of 2D turbulence for
the Euler-Poisson  system \cite{prl}, we have derived a kinetic equation of
the form
\begin{eqnarray}
\label{E47} {\partial \overline{\omega}\over\partial
t}+\overline{\bf u}\nabla\overline{\omega}={\epsilon^{2}\tau\over 8\pi^{2}}{\partial\over\partial r^{\mu}}\int d^{2}{\bf r}'
K^{\mu\nu}\biggl\lbrace \overline{\omega}'(\sigma_{0}-\overline{\omega}'){\partial\overline{\omega}\over\partial r^{\nu}}-\overline{\omega}(\sigma_{0}-\overline{\omega}){\partial\overline{\omega}'\over\partial
r'^{\nu}}\biggr\rbrace,
\end{eqnarray}
where $K^{\mu\nu}$ is the tensor
\begin{eqnarray}
\label{E48} K^{\mu\nu}= {1\over \xi^{2}}\biggl
(\delta^{\mu\nu}-{\xi^{\mu}\xi^{\nu}\over \xi^{2}}\biggr ),
\end{eqnarray}
and ${\mb\xi}={\bf r}'-{\bf r}$.  This equation is valid in the
two-levels approximation of the statistical theory $\omega=\lbrace
\sigma_{0},0\rbrace$. Once again, the multi-levels case would lead
to $N$ coupled differential equations for $\rho({\bf
r},\sigma,t)$, the density probability of finding the vorticity
level $\omega=\sigma$ in ${\bf r}$ at time $t$.  Alternatively, if
we implement a closure approximation of the form
$\omega_{2}=1/C''(\overline{\omega})$, similar to Eq. (\ref{c1}),
we obtain a simpler equation 
\begin{eqnarray}
\label{E49} {\partial \overline{\omega}\over\partial
t}+\overline{\bf u}\nabla\overline{\omega}={\epsilon^{2}\tau\over 8\pi^{2}}{\partial\over\partial r^{\mu}}\int d^{2}{\bf r}'
K^{\mu\nu}\biggl\lbrace {1\over C''(\overline{\omega}')}{\partial\overline{\omega}\over\partial r^{\nu}}-{1\over C''(\overline{\omega})}{\partial\overline{\omega}'\over\partial
r'^{\nu}}\biggr\rbrace,
\end{eqnarray}
which takes into account several features of the process of ``inviscid
relaxation" in 2D turbulence. Unfortunately, as discussed in
\cite{prl}, Eqs. (\ref{E47}) and (\ref{E49}) do {\it not} conserve
energy. Non-markovian effects may be necessary to restore the
conservation of energy.

By using a Maximum Entropy Production Principe (MEPP), Robert \& Sommeria
\cite{rsmepp} have proposed a relaxation equation for the local distribution of vorticity $\rho({\bf r},\sigma,t)$ of the form
\begin{eqnarray}
\label{wq1}
{\partial\rho\over\partial t}+\overline{\bf u}\nabla\rho=\nabla\biggl\lbrace D\biggl\lbrack \nabla\rho+\beta\rho (\sigma-\rho)\nabla\psi\biggr \rbrack\biggr\rbrace.
\end{eqnarray}
From this equation, we can deduce a hierarchy of equations for the moments $\overline{\omega^{n}}=\int \rho\sigma^{n} d\sigma$. For the first moment $\overline{\omega}$, we have
\begin{eqnarray}
\label{wq2}
{\partial\overline{\omega}\over\partial t}+\overline{\bf u}\nabla\overline{\omega}=\nabla\biggl\lbrace D\biggl\lbrack \nabla\overline{\omega}+\beta\omega_{2}\nabla\psi\biggr \rbrack\biggr\rbrace.
\end{eqnarray}
Kazantsev {\it et al.} \cite{kazantsev} proposed to close the hierarchy of equations
with a Gaussian approximation. The resulting equation converges to a state of minimum
enstrophy. More generally, we propose to
close the hierarchy of equations by the relation
$\omega_{2}=1/C''(\overline{\omega})$ so that
\begin{eqnarray}
\label{wd3}
{\partial\overline{\omega}\over\partial t}+\overline{\bf u}\nabla\overline{\omega}=\nabla\biggl\lbrace D\biggl\lbrack \nabla\overline{\omega}+{\beta\over C''(\overline{\omega})}\nabla\psi\biggr \rbrack\biggr\rbrace.
\end{eqnarray}
The function $C(\overline{\omega})$ is a free function which has to be
adapted to the context (see Sec. \ref{sec_heuristic}). This
indetermination reflects the fundamental observation that there is
{\it no universal} form of entropy $S[\overline{\omega}]$ in 2D
hydrodynamics \cite{gt}. The enstrophy \cite{leith}, the Fermi-Dirac
entropy \cite{staquet}, the Boltzmann entropy \cite{jm} (leading to a
sinh-Poisson equation) or the functional proposed by Ellis {\it et
al.} \cite{ellis} (leading to a Gamma law for the vorticity
fluctuations) are {\it particular} functionals $S[\overline{\omega}]$
which prove to be more relevant than others in a {\it specific} context
(geophysical flows
\cite{kazantsev}, Jupiter's Great Red Spot \cite{bs}, 2D
turbulence \cite{montgomery} and jovian atmosphere \cite{ellis}
respectively). None of them has a universal domain of validity but
generalized entropies can possibly be regrouped in ``classes of
equivalence'' \cite{gt}. Tsallis entropies, sometimes arising in 2D
hydrodynamics \cite{boghosian}, are special because they are due to
{\it incomplete relaxation}. They are particular H-functions
associated to a dynamical equilibrium, not to a thermodynamical
equilibrium (they are not true entropies). The self-confinement of 2D
vortices can be explained by a lack of mixing
\cite{brands}. It can be taken into account in the statistical theory
by using relaxation equations with a space-dependant diffusion
coefficient related to the local fluctuations of the vorticity
\cite{rr}. The same arguments can be invoked to account for the
confinement of galaxies in astrophysics \cite{csr}.

In conclusion, the generalized kinetic equations introduced previously
can provide a simple parametrization of turbulence
(mixing) in the context of the violent relaxation of stellar systems
and 2D vortices. They can take into account incomplete relaxation thanks
to a varying diffusion coefficient
\cite{rr,csr}. In addition, due to the {\it thermodynamical analogy}
discussed in Sec. \ref{sec_meta}, they can also serve as powerful
numerical algorithms to compute nonlinearly dynamically stable
solutions of the Vlasov-Poisson or 2D Euler-Poisson systems. This is a
great practical interest of these equations independently of the
statistical theory.

\section{Properties of the generalized Landau equation}
\label{sec_prop}

\subsection{Conservation laws}
\label{sec_cons}

In this section, we derive the conservation laws satisfied by the
generalized Landau equation (\ref{wd12}).  Let us first introduce
the current of diffusion
\begin{eqnarray}
\label{cons1}
J_{f}^{\mu}= -\int d^{3}{\bf v}_{1}K^{\mu\nu} \biggl\lbrace g(f_{1})h(f) {\partial f\over\partial v^{\nu}}-g(f)h(f_{1}){\partial f_{1}\over\partial v_{1}^{\nu}}\biggr\rbrace.
\end{eqnarray}
The time variation of energy can be written
\begin{equation}
\label{cons2} \dot E=\int {\partial f\over\partial t}\biggl ({v^{2}\over
2}+\Phi\biggr )d^{3}{\bf r} d^{3}{\bf v}=\int {\bf J}_{f}\cdot {\bf v} \ d^{3}{\bf r}d^{3}{\bf v},
\end{equation}
where we have used an integration by parts.
Introducing the current (\ref{cons1}) in Eq. (\ref{cons2}), we obtain
\begin{eqnarray}
\label{cons3}
\dot E= -\int d^{3}{\bf r} d^{3}{\bf v}d^{3}{\bf v}_{1} K^{\mu\nu} v^{\mu} \biggl\lbrace g(f_{1})h(f) {\partial f\over\partial v^{\nu}}-g(f)h(f_{1}){\partial f_{1}\over\partial v_{1}^{\nu}}\biggr\rbrace.
\end{eqnarray}
Interchanging the dummy variables  ${\bf v}$ and ${\bf v}_1$, we
get
\begin{eqnarray}
\label{cons4}
\dot E= \int d^{3}{\bf r} d^{3}{\bf v}d^{3}{\bf v}_{1} K^{\mu\nu} v_{1}^{\mu} \biggl\lbrace g(f_{1})h(f) {\partial f\over\partial v^{\nu}}-g(f)h(f_{1}){\partial f_{1}\over\partial v_{1}^{\nu}}\biggr\rbrace,
\end{eqnarray}
where we have exploited the symmetrical form of the diffusion
current. Taking the half sum of the last two expressions, we find
\begin{eqnarray}
\label{cons5}
\dot E={1\over 2} \int d^{3}{\bf r}d^{3}{\bf v}d^{3}{\bf v}_{1} K^{\mu\nu} (v_{1}^{\mu}-v^{\mu}) \biggl\lbrace g(f_{1})h(f) {\partial f\over\partial v^{\nu}}-g(f)h(f_{1}){\partial f_{1}\over\partial v_{1}^{\nu}}\biggr\rbrace.
\end{eqnarray}
Noting that
\begin{eqnarray}
\label{cons6}
K^{\mu\nu}u^{\nu}=0,
\end{eqnarray}
according to Eq. (\ref{wd13}), we finally establish that $\dot
E=0$. Therefore, the generalized Landau equation conserves energy. We
can show by a similar procedure that it also conserves angular
momentum and impulse.

\subsection{Generalized H-theorem}
\label{sec_h}

The rate of production of generalized entropy  is
given by
\begin{eqnarray}
\label{h1}
\dot S=-\int C'(f){\partial f\over\partial t}d^{3}{\bf r}d^{3}{\bf v}=-\int C''(f)J^{\mu}{\partial f\over\partial v^{\mu}}d^{3}{\bf r}d^{3}{\bf v}=-\int {h(f)\over g(f)}J^{\mu}{\partial f\over\partial v^{\mu}}d^{3}{\bf r} d^{3}{\bf v},
\end{eqnarray}
where we have used Eq. (\ref{ge5}) to get the last equality.
Inserting the current (\ref{cons1}) in Eq. (\ref{h1}), we obtain
\begin{eqnarray}
\label{hh2}
\dot S= \int d^{3}{\bf r}d^{3}{\bf v}d^{3}{\bf v}_{1} {h\over g}{\partial f\over\partial v^{\mu}} K^{\mu\nu}  \biggl\lbrace g_{1}h {\partial f\over\partial v^{\nu}}-g h_{1}{\partial f_{1}\over\partial v_{1}^{\nu}}\biggr\rbrace.
\end{eqnarray}
Interchanging the dummy variables  ${\bf v}$ and ${\bf v}_1$, we
get
\begin{eqnarray}
\label{hh3}
\dot S=- \int d^{3}{\bf r} d^{3}{\bf v}d^{3}{\bf v}_{1} {h_{1}\over g_{1}}{\partial f_{1}\over\partial v_{1}^{\mu}} K^{\mu\nu} \biggl\lbrace g_{1}h {\partial f\over\partial v^{\nu}}-g h_{1}{\partial f_{1}\over\partial v_{1}^{\nu}}\biggr\rbrace.
\end{eqnarray}
Taking the half sum of the foregoing equations we find that
\begin{eqnarray}
\label{hh4}
\dot S={1\over 2} \int d^{3}{\bf r}d^{3}{\bf v}d^{3}{\bf v}_{1}{1\over g g_{1}}\biggl\lbrace g_{1} h{\partial f\over\partial v^{\mu}}-g h_{1}{\partial f_{1}\over\partial v_{1}^{\mu}}\biggr\rbrace K^{\mu\nu} \biggl\lbrace g_{1}h {\partial f\over\partial v^{\nu}}-g h_{1}{\partial f_{1}\over\partial v_{1}^{\nu}}\biggr\rbrace.
\end{eqnarray}
Noting that $X^{\mu}K^{\mu\nu}X^{\nu}=X^{2}-({\bf X}\cdot{\bf
u})^{2}/u^{2}\ge 0$, we conclude that $\dot S\ge 0$. Therefore,
the generalized Landau equation satisfies a generalized
$H$-theorem.

\subsection{Equilibrium distribution}
\label{sec_ed}

Taking the derivative of Eq. (\ref{ge3}) with respect to ${\bf v}$, we get
\begin{eqnarray}
\label{ed1}
C''(f_{eq}){\partial f_{eq}\over\partial {\bf v}}=-\beta ({\bf v}-{\bf \Omega}{\times} {\bf r}-{\bf U}).
\end{eqnarray}
Inserting this relation in the current (\ref{cons1}) and using
Eq. (\ref{ge5}), we obtain
\begin{eqnarray}
\label{m6}
J_{f}^{\mu}=-\beta \int d^{3}{\bf v}_{1}K^{\mu\nu}g g_{1}
(v_{1}^{\nu}-v^{\nu}),
\end{eqnarray}
which vanishes identically in virtue of Eq. (\ref{cons6}). Therefore,
an extremum of the generalized entropy at fixed mass, energy, angular
momentum and impulse is a stationary solution of the generalized
Landau equation. Alternatively, a stationary solution satisfies $\dot
S=0$ hence ${\bf J}_{f}={\bf 0}$. We can show that this condition
implies that $f_{eq}$ is of the form (\ref{ge3}). The proof is similar
to the one given for the ordinary Landau equation.

Finally, considering the linear stability of a stationary solution of the generalized Landau equation, we can derive the general relation \cite{gt}
\begin{eqnarray}
\label{m7}
2\lambda\delta^{2}J=\delta^{2}\dot S\ge 0,
\end{eqnarray}
connecting the growth rate $\lambda$ of the perturbation (such
that $\delta f\sim e^{\lambda t}$) to the second order variations
of the free energy $J=S-\beta E$ and the second order variations
of the rate of entropy production $\delta^{2}\dot S\ge 0$. This
aesthetic relation implies that a stationary solution of the
generalized Landau equation is linearly stable if, and only if, it
is an entropy {\it maximum} at fixed mass and energy \cite{gt}.

\section{The thermal bath approximation}
\label{sec_tb}

The generalized Landau equation (\ref{wd12}) can be put in a form reminiscent
of a Fokker-Planck  equation
\begin{eqnarray}
\label{tb1} {df\over dt}={\partial\over\partial
v^{\mu}}\biggl\lbrack D^{\mu\nu}h(f) {\partial f\over\partial
v^{\nu}}+g(f)\eta^{\mu}\biggr \rbrack,
\end{eqnarray}
by introducing a diffusion tensor
\begin{eqnarray}
\label{tb2}
D^{\mu\nu}=\int K^{\mu\nu} g(f_{1})d^{3}{\bf v}_{1},
\end{eqnarray}
and a friction term
\begin{eqnarray}
\label{tb3}
\eta^{\mu}=-\int K^{\mu\nu}h(f_{1}){\partial f_{1}\over\partial v_{1}^{\nu}}d^{3}{\bf v}_{1}.
\end{eqnarray}
The ordinary Fokker-Planck equation \cite{risken} can be written
\begin{eqnarray}
\label{tb4} {df\over dt}={\partial^{2}\over\partial v^{\mu}\partial v^{\nu}}(f\zeta^{\mu\nu})+{\partial \over\partial v^{\mu}}(f\zeta^{\mu}),
\end{eqnarray}
where
\begin{eqnarray}
\label{tb5}\zeta^{\mu\nu}= {1\over 2}{\langle \Delta v^{\mu}\Delta v^{\nu}\rangle\over \Delta t}, \qquad \zeta^{\mu}=-{\langle \Delta v^{\mu}\rangle\over
\Delta t},
\end{eqnarray}
are the first (friction) and second (diffusion) moments of the
velocity deviation. An equivalent form of the Fokker-Planck equation is
\begin{eqnarray}
\label{tb6} {df\over dt}={\partial\over\partial v^{\mu}}\biggl\lbrack \zeta^{\mu\nu}{\partial f\over\partial v^{\nu}}+\biggl ({\partial\zeta^{\mu\nu}\over\partial v^{\nu}}+\zeta^{\mu}\biggr )f\biggr\rbrack.
\end{eqnarray}

For the ordinary Landau equation, we have $g(f)=f$ et $h(f)=1$. Therefore, Eq. (\ref{tb1}) becomes
\begin{eqnarray}
\label{tb7} {df\over dt}={\partial\over\partial
v^{\mu}}\biggl\lbrack D^{\mu\nu} {\partial f\over\partial
v^{\nu}}+f\eta^{\mu}\biggr \rbrack.
\end{eqnarray}
By comparing with the Fokker-Planck equation (\ref{tb6}),
we obtain the classical relations
\begin{eqnarray}
\label{tb8} \zeta^{\mu\nu}=D^{\mu\nu}, \qquad {\partial \zeta^{\mu\nu}\over\partial v^{\nu}}+\zeta^{\mu}=\eta^{\mu}.
\end{eqnarray}
According to Eq. (\ref{tb8}), there is a difference between ${\mb
\eta}$ and $\langle \Delta {\bf v}\rangle$ due to the velocity
dependance of the diffusion tensor $D^{\mu\nu}$. Noting that
\begin{eqnarray}
\label{tb9}
{\partial D^{\mu\nu}\over \partial v^{\nu}}=\int {\partial K^{\mu\nu}\over\partial v^{\nu}}f_{1}d^{3}{\bf v}_{1}=-\int {\partial K^{\mu\nu}\over\partial v_{1}^{\nu}} f_{1}d^{3}{\bf v}_{1}=\int K^{\mu\nu}{\partial f_{1}\over\partial v_{1}^{\nu}}d^{3}{\bf v}_{1}=-\eta^{\mu},
\end{eqnarray}
we find that Eq. (\ref{tb8}) yields
\begin{eqnarray}
\label{tb10}
{\mb\eta}=-{1\over 2}{\langle \Delta {\bf
v}\rangle\over\Delta t}.
\end{eqnarray}
Therefore, in the classical case, the
vector $-{\mb \eta}$ represents {\it half} the friction force.

The generalized Landau equation (\ref{tb1}) can be rewritten
\begin{eqnarray}
\label{tb11} {df\over dt}={\partial^{2}\over\partial
v^{\mu}\partial v^{\nu}}( D^{\mu\nu}{H(f)} )+ {\partial \over\partial
v^{\nu}}\biggl ( {g(f)}\eta^{\mu}-{H(f)}{\partial D^{\mu\nu}\over\partial v^{\nu}}\biggr ),
\end{eqnarray}
where $H$ is a primitive of $h$. By comparing with the {\it ordinary} Fokker-Planck equation (\ref{tb4}), we get
\begin{eqnarray}
\label{tb6bis} \zeta^{\mu\nu}=D^{\mu\nu}{H(f)\over f},\qquad  \zeta^{\mu}={g(f)\over f}\eta^{\mu}-{H(f)\over f}{\partial D^{\mu\nu}\over\partial v^{\nu}}.
\end{eqnarray}
If we impose the relations $g(f)=f$ and $h(f)=f C''(f)$ of Sec. \ref{sec_s}, Eq. (\ref{tb6bis}) can be rewritten
\begin{eqnarray}
\label{tb12}
{1\over 2}{\langle \Delta v^{\mu}\Delta v^{\nu}\rangle\over
\Delta t}=D^{\mu\nu} f \biggl \lbrack {C(f)\over f}\biggr \rbrack'.
\end{eqnarray}
This relation was noted in \cite{gt}. In the particular case of the
Tsallis entropy, Eq. (\ref{tb12}) reduces to the form considered by Borland
\cite{borland}.  In the Borland approach, generalized Fokker-Planck
equations arise because the transition probabilities $\zeta^{\mu\nu}$
and $\zeta^{\mu}$ depend explicitly on the distribution
function. A different  approach is followed by Kaniadakis
\cite{kaniadakis}. Starting from a kinetic interaction principle
(KIP), he obtains a generalized Fokker-Planck equation of the form
\begin{eqnarray}
\label{tb13} {df\over dt}={\partial\over\partial v^{\mu}}\biggl\lbrack h(f) \zeta^{\mu\nu}{\partial f\over\partial v^{\nu}}+g(f)\biggl ({\partial\zeta^{\mu\nu}\over\partial v^{\nu}}+\zeta^{\mu}\biggr )f\biggr\rbrack.
\end{eqnarray}
This corresponds to his Eq. (17) in our notations. Comparing with our
equation (\ref{tb1}), we find that the classical relations (\ref{tb8})
are preserved in this generalized framework. This is much more
aesthetic than Eq. (\ref{tb6bis}). This is also more physical because we
would expect that the transition moments depend only on the form of
interaction, not explicitly on the distribution function. Besides, the
other approach leads to ambiguity because, comparing Eqs. (\ref{tb1}) and (\ref{tb6}),
we get
\begin{eqnarray}
\label{tb14} \zeta^{\mu\nu}=D^{\mu\nu}h(f),\qquad  {\partial \zeta^{\mu\nu}\over\partial v^{\nu}}+\zeta^{\mu}={g(f)\over f}\eta^{\mu},
\end{eqnarray}
which differ from Eq. (\ref{tb6bis}) except in the classical case.

Physically, the Landau equation describes the statistical evolution of
a system of self-interacting particles in a mean-field
approximation. This description treats all the particles on the same
footing and conserves the energy of the whole system.  This
corresponds to a microcanonical description. We can alteratively view
the Landau equation as describing the evolution of a test particle
(described by $f({\bf v},t)$) interacting with field particles
(described by $f({\bf v}_{1},t)$). In the so-called {\it thermal bath
approximation} (canonical description), we consider that the field
particles are in statistical equilibrium and replace $f({\bf
v}_{1},t)$ by their equilibrium distribution $f_{eq}({\bf v}_{1})$
given by Eq. (\ref{ge3}). This approximation transforms an
integro-differential equation (Landau) into a differential equation
(Kramers). Combining Eqs. (\ref{tb3}), (\ref{ed1}) and (\ref{ge5}), we
get
\begin{eqnarray}
\label{tb15} \eta^{\mu}=\beta\int K^{\mu\nu} g(f_{1}) w_{1}^{\nu}
d^{3}{\bf v}_{1},
\end{eqnarray}
with ${\bf w}={\bf v}-{\bf\Omega}{\times} {\bf r}-{\bf U}$. Now,
using Eqs. (\ref{cons6}) and (\ref{tb2}), we have equivalently
\begin{eqnarray}
\label{tb16} \eta^{\mu}=\beta\int K^{\mu\nu} g(f_{1}) w^{\nu}
d^{3}{\bf v}_{1}  =\beta D^{\mu\nu} w^{\nu}.
\end{eqnarray}
This relation can be interpreted as a generalized Einstein
relation. In the test particle approach, the kinetic equation
(\ref{tb1}) thus becomes
\begin{eqnarray}
\label{tb17} { df\over dt}={\partial\over\partial
v^{\mu}}\biggl \lbrace D^{\mu\nu}\biggl\lbrack h(f){\partial
f\over\partial v^{\nu}}+\beta g(f) w^{\nu}\biggr \rbrack\biggr
\rbrace.
\end{eqnarray}
This equation will be called the generalized Kramers equation.
Assuming for simplicity that the diffusion is isotropic, taking
${\mb\Omega}={\bf U}={\bf 0}$ and imposing the relations of
Eq. (\ref{s1}), we recover the generalized Kramers equation
\begin{eqnarray}
\label{tb18} {df\over dt}={\partial\over\partial
{\bf v}}\biggl \lbrace D\biggl\lbrack fC''(f){\partial
f\over\partial {\bf v}}+\beta f {\bf v}\biggr \rbrack\biggr
\rbrace,
\end{eqnarray}
proposed in \cite{gt}. In the present approach, the generalized
Kramers equation is derived from the generalized Landau equation by a
systematic procedure. This makes possible to determine explicitly the
expression of the diffusion coefficient (see
Secs. \ref{sec_diff}-\ref{sec_exp}) which was left unspecified by the
MEPP approach \cite{gt}.

We recall also that in the long time limit (or strong friction
limit $\xi=D\beta\rightarrow +\infty$), the generalized Kramers
equation leads to the generalized Smoluchowski equation 
\begin{eqnarray}
\label{tb19}
{\partial\rho\over\partial t}=\nabla\Bigl\lbrack {1\over\xi}(\nabla p+\rho\nabla\Phi)\Bigr\rbrack,
\end{eqnarray}
where $p(\rho)$ has the interpretation of a pressure \cite{gt}. The
generalized Smoluchowski equation was first introduced in
\cite{csr}. The gravitational (i.e.
attracting) Smoluchowski-Poisson and generalized
Smoluchowski-Poisson systems have been studied in \cite{sp,gsp}
(see also \cite{biler} for a connexion with mathematical results).
We shall say that a kinetic equation has a generalized canonical
thermodynamical structure if it increases the generalized free
energy $J=S-\beta E$ at fixed inverse temperature $\beta$ and mass
$M$ (the function $J$ differs from the usual free energy $F=E-TS$
by a factor $-\beta$). We have shown that the generalized
Kramers equation and the generalized Smoluchowski equation possess
such a thermodynamical structure \cite{gt}. We can formally obtain
a microcanonical formulation of the generalized Kramers and
Smoluchowski equations by letting the inverse temperature $\beta$
depend on time, i.e. $\beta=\beta(t)$, so as to conserve energy.
With this modification, the generalized Kramers and Smoluchowski
equations have a microcanonical thermodynamical structure
\cite{gt}.

\section{Diffusion coefficient and dynamical friction}
\label{sec_diff}

\subsection{Generalized Rosenbluth potentials}
\label{sec_rose}

In order to evaluate the expressions of diffusion coefficient and
 friction, it is convenient to introduce auxiliary functions which
are called the Rosenbluth potentials \cite{bt}. Noting that
\begin{equation}
{\partial^{2} u\over\partial v^{\mu}\partial v^{\nu}}= {u^{2}\delta^{\mu\nu}-u^{\mu}u^{\nu}\over u^{3}},
\label{idd}
\end{equation}
we can rewrite the diffusion tensor (\ref{tb2}) in the form
\begin{equation}
D^{\mu\nu}=A {\partial^{2} \chi\over\partial v^{\mu}\partial
v^{\nu}}({\bf v}), \label{diff2}
\end{equation}
where
\begin{equation}
\chi({\bf v})=\int g(f_{1}) |{\bf v}-{\bf v}_{1}| d^{3}{\bf v}_{1},
\label{g}
\end{equation}
is the generalized Rosenbluth potential  associated to  the
diffusion. Writing the friction term (\ref{tb3}) in the form
\begin{eqnarray}
\label{m17} \eta^{\mu}=-\int K^{\mu\nu}{\partial
H(f_{1})\over\partial v_{1}^{\nu}}d^{3}{\bf v}_{1},
\end{eqnarray}
and integrating by parts, we obtain
\begin{eqnarray}
\eta^{\mu}=\int  {\partial K^{\mu\nu}\over\partial v_{1}^{\nu}} H(f_{1}) d^{3}{\bf v}_{1}.
\label{friction2}
\end{eqnarray}
Noting that
\begin{eqnarray}
{\partial K^{\mu\nu}\over\partial v_{1}^{\mu}}=2A  {u^{\nu}\over u^{3}}=-2A  {\partial \over\partial v^{\mu}}\biggl ({1\over u}\biggr ),
\label{idd2}
\end{eqnarray}
we can rewrite the friction term as
\begin{equation}
\eta^{\mu}=-2A{\partial \lambda\over\partial v^{\mu}}({\bf v}),
\label{fric3}
\end{equation}
where
\begin{equation}
\lambda({\bf v})=\int {H(f_{1})\over |{\bf v}-{\bf
v}_{1}|}d^{3}{\bf v}_{1},
\label{h}
\end{equation}
is the generalized Rosenbluth potential associated to the friction
(we need to impose that $H(f_{1})$ tends to zero for $|{\bf
v}_{1}|\rightarrow +\infty$ to make the integral well defined;
this fixes the constant of integration in $H$). We note also that
\begin{eqnarray}
\label{frw1}{\partial D^{\mu\nu}\over \partial v^{\nu}}=\int  {\partial K^{\mu\nu}\over\partial v^{\nu}} g(f_{1}) d^{3}{\bf v}_{1}=- \int {\partial K^{\mu\nu}\over\partial v_{1}^{\nu}} g(f_{1}) d^{3}{\bf v}_{1}=2A{\partial \sigma\over\partial v^{\mu}}({\bf v}),
\end{eqnarray}
where
\begin{equation}
\label{frw2}
\sigma({\bf v})=\int {g(f_{1})\over |{\bf v}-{\bf
v}_{1}|}d^{3}{\bf v}_{1},
\end{equation}
is the potential associated to the velocity dependance of the
diffusion coefficient. In the usual thermodynamical framework
where $g(f)=H(f)=f$, the potentials $\lambda$ and $\sigma$
coincide.

\subsection{Isotropic distribution of velocities}
\label{sec_iso}

When the velocity distribution of the field particles is
isotropic, i.e. $f_{1}=f(v_{1})$, we can obtain more explicit
expressions for the Rosenbluth potentials. For reasons of
symmetry, the Rosenbluth potentials depend only on $v=|{\bf v}|$.
To determine $\chi(v)$ and $\lambda(v)$, we use the identity
\begin{equation}
{1\over |{\bf v}-{\bf v}_{1}|}=\sum_{l=0}^{+\infty}{v_{<}^{l}\over v_{>}^{l+1}}P_{l}(\cos\gamma),
\label{legendre}
\end{equation}
where $v_{<}$ and $v_{>}$ denote the smallest and largest value of $v$ and $v_{1}$, $P_{l}(x)$ is a Legendre polynomial and $\gamma$ is the angle between ${\bf v}$ and ${\bf v}_{1}$. Then, we have
\begin{equation}
\lambda(v)=2\pi\sum_{l=0}^{+\infty}\int_{0}^{+\infty}{v_{1}^{2}v_{<}^{l}\over v_{>}^{l+1}}H\lbrack f(v_{1})\rbrack dv_{1}\int_{0}^{\pi}P_{l}(\cos\gamma)\sin\gamma d\gamma.
\label{h2}
\end{equation}
With the change of variable $x=\cos\gamma$, we get
\begin{equation}
\lambda(v)=2\pi\sum_{l=0}^{+\infty}\int_{0}^{+\infty}{v_{1}^{2}v_{<}^{l}\over v_{>}^{l+1}}H\lbrack f(v_{1})\rbrack dv_{1}\int_{-1}^{+1}P_{l}(x)dx.
\label{h3}
\end{equation}
Using the identity  $\int_{-1}^{+1}P_{l}(x)dx=2\delta_{l0}$, the foregoing expression can be simplified in
\begin{equation}
\lambda(v)=4\pi\int_{0}^{+\infty}{v_{1}^{2}\over v_{>}}H\lbrack f(v_{1})\rbrack dv_{1}=4\pi\biggl \lbrack {1\over v}\int_{0}^{v}v_{1}^{2}H\lbrack f(v_{1})\rbrack dv_{1}+\int_{v}^{+\infty}v_{1}H\lbrack f(v_{1})\rbrack dv_{1}\biggr\rbrack.
\label{h4}
\end{equation}
Similarly, we have
\begin{equation}
\sigma(v)=4\pi\biggl \lbrack {1\over
v}\int_{0}^{v}v_{1}^{2}g\lbrack f(v_{1})\rbrack
dv_{1}+\int_{v}^{+\infty}v_{1}g\lbrack f(v_{1})\rbrack
dv_{1}\biggr\rbrack. \label{h4b}
\end{equation}

To determine $\chi(v)$, we write $|{\bf v}-{\bf v}_{1}|=(v^{2}+v_{1}^{2}-2vv_{1}\cos\gamma)/|{\bf v}-{\bf v}_{1}|$ and we use the identity (\ref{legendre}). We then obtain
\begin{equation}
\chi(v)=2\pi\sum_{l=0}^{+\infty}\int_{0}^{+\infty}{v_{1}^{2}v_{<}^{l}\over v_{>}^{l+1}}g\lbrack f(v_{1})\rbrack dv_{1}\int_{0}^{\pi}(v^{2}+v_{1}^{2}-2vv_{1}\cos\gamma)   P_{l}(\cos\gamma)\sin\gamma d\gamma.
\label{g2}
\end{equation}
With the change of variables $x=\cos\gamma$, we get
\begin{equation}
\chi(v)=2\pi\sum_{l=0}^{+\infty}\int_{0}^{+\infty}{v_{1}^{2}v_{<}^{l}\over v_{>}^{l+1}}g\lbrack f(v_{1})\rbrack dv_{1}\int_{-1}^{+1}(v^{2}+v_{1}^{2}-2vv_{1}x)  P_{l}(x)dx.
\label{g3}
\end{equation}
Using the identity $\int_{-1}^{+1}x P_{l}(x)dx={2\over 3}\delta_{l1}$, we obtain after some manipulations
\begin{equation}
\chi(v)={4\pi v\over 3}\biggl\lbrack \int_{0}^{v}\biggl (3v_{1}^{2}+{v_{1}^{4}\over v^{2}}\biggr )g\lbrack f(v_{1})\rbrack dv_{1}+\int_{v}^{+\infty}\biggl ({3 v_{1}^{3}\over v}+vv_{1}\biggr )g\lbrack f(v_{1})\rbrack dv_{1}\biggr\rbrack.
\label{g4}
\end{equation}

\subsection{The  diffusion coefficient}
\label{sec_dc}

We are now in a position to determine an explicit expression for
the diffusion tensor $D^{\mu\nu}$ that is valid for an arbitrary
isotropic distribution function of the field particles. Starting
from the identity
\begin{equation}
{\partial^{2}\chi\over\partial v^{\mu}\partial v^{\nu}}={v^{\mu}v^{\nu}\over v^{2}}\biggl ({d^{2}\chi\over dv^{2}}-{1\over v}{d\chi\over dv}\biggr )+{1\over v}{d\chi\over dv}\delta^{\mu\nu},
\label{g5}
\end{equation}
the diffusion coefficient (\ref{diff2}) can be put in the form
\begin{equation}
D^{\mu\nu}=\biggl (D_{||}-{1\over 2}D_{\perp}\biggr ){v^{\mu}v^{\nu}\over v^{2}}+{1\over 2}D_{\perp}\delta^{\mu\nu},
\label{diff3}
\end{equation}
where
\begin{equation}
D_{\perp}=2A {1\over v}{d\chi\over dv},
 \label{Dperp}
\end{equation}
and
\begin{equation} D_{||}=A {d^{2}\chi \over dv^{2}}, \label{Dpar}
\end{equation}
are the diffusion coefficients in the directions perpendicular and
parallel to the velocity of the test particle. To see that, we
 consider a system of coordinates where the $z$-axis is
taken in the direction of ${\bf v}$ so that $v_{x}=v_{y}=0$ and
$v_{z}=v$. In this system of coordinates, all the non-diagonal
elements of $D^{\mu\nu}$ vanish while $D_{xx}=D_{yy}={1\over
2}D_{\perp}$ and $D_{zz}=D_{||}$. According to formula (\ref{g4}),
we have explicitly
\begin{equation}
D_{\perp}={8\pi\over 3}A {1\over v}\biggl\lbrack \int_{0}^{v}\biggl (3v_{1}^{2}-{v_{1}^{4}\over v^{2}}\biggr )g\lbrack f(v_{1})\rbrack dv_{1}+2v\int_{v}^{+\infty}v_{1} g\lbrack f(v_{1})\rbrack dv_{1}\biggr\rbrack,
\label{Dperp2}
\end{equation}
\begin{equation}
D_{||}={8\pi\over 3}A {1\over v}\biggl\lbrack \int_{0}^{v}{v_{1}^{4}\over v^{2}} g\lbrack f(v_{1})\rbrack dv_{1}+v\int_{v}^{+\infty}v_{1}g\lbrack f(v_{1})\rbrack dv_{1}\biggr\rbrack.
\label{Dpar2}
\end{equation}

\subsection{The dynamical friction}
\label{sec_df}

The friction term (\ref{fric3}) can be simplified similarly. For an isotropic
velocity distribution,
\begin{equation}
{\partial \lambda\over\partial v^{\mu}}={d\lambda \over dv}{v^{\mu}\over v},
\label{h5}
\end{equation}
so that
\begin{equation}
{\mb \eta}=-2A {1\over v}{d\lambda\over dv}{\bf v}.
\label{Fric1}
\end{equation}
Using Eq. (\ref{h4}), we get
\begin{equation}
{\mb\eta}=8\pi A {{\bf v}\over v^{3}}\int_{0}^{v}v_{1}^{2}H\lbrack f(v_{1})\rbrack dv_{1}.
\label{Fric2}
\end{equation}
Similarly, we have
\begin{eqnarray}
\label{m15}{\partial D^{\mu\nu}\over \partial v^{\nu}}=-8\pi A {{\bf v}\over v^{3}}\int_{0}^{v}v_{1}^{2}g\lbrack f(v_{1})\rbrack dv_{1}.
\end{eqnarray}
We note that when the field particles have an isotropic velocity distribution, the dynamical friction experienced by the test particle
\begin{eqnarray}
\label{m15b}\langle {\bf F}\rangle_{friction}=-8\pi A {{\bf v}\over v^{3}}\int_{0}^{v}v_{1}^{2} \biggl\lbrack {g(f)\over f}H(f_{1})+{H(f)\over f}g(f_{1})\biggr \rbrack dv_{1},
\end{eqnarray}
 is parallel and
opposite to its velocity ${\bf v}$. Moreover,
it is due only to field particles  with velocity $v_{1}<v$. In
fact, we can obtain this result without calculation. Indeed,
according to Eqs. (\ref{fric3}) et (\ref{h}), we have
\begin{equation}
{\mb\eta}=2A \int {{\bf v}-{\bf v}_{1}\over |{\bf v}-{\bf v}_{1}|^{3}}H\lbrack f_{1}\rbrack d^{3}{\bf v}_{1},
\label{Fric3}
\end{equation}
and a similar expression for ${\partial D^{\mu\nu}/ \partial
v^{\nu}}$. Now, Eq. (\ref{Fric3})  is analogous to the
gravitational force created in ${\bf v}$ by a distribution of mass
with density $H\lbrack f(v_{1})\rbrack$ where ${\bf v}$ plays the
role of the position ${\bf r}$. According to Newton's law, the
gravitational force created in ${\bf r}$ by an isotropic
distribution of mass depends only on the mass interior to $r$ and
is given by an expression equivalent to Eq. (\ref{Fric2}).

\section{Explicit results for typical distribution functions}
\label{sec_exp}

\subsection{Isothermal distributions}
\label{sec_it}

We shall now obtain explicit expressions of diffusion coefficient
and friction force for typical distribution functions. First, we
consider the case of ordinary thermodynamics based on the
Boltzmann entropy. In the thermal bath approximation, the field
stars are at statistical equilibrium with the isothermal
distribution
\begin{equation}
f_{eq}({\bf v}_{1})=\rho\biggl ({\beta \over 2\pi}\biggr
)^{3/2}e^{-\beta  {v_{1}^{2}\over 2}}.
 \label{it1}
\end{equation}
For the Boltzmann entropy $C(f)=f\ln f$, we have $g(f)=H(f)=f$.
The diffusion coefficient and the friction coefficient can then be
calculated explicitly by substituting Eq. (\ref{it1}) in Eqs.
(\ref{Dperp2}), (\ref{Dpar2}) and (\ref{Fric2}), and performing
the integrals.  Introducing the notation $X=\sqrt{\beta v^{2}/2}$
and the function
\begin{equation}
G(X)={2\over\sqrt{\pi}}{1\over X^{2}}\int_{0}^{X}t^{2}e^{-t^{2}}dt={1\over 2 X^{2}}\biggl\lbrack {\rm erf}(X)-{2X\over \sqrt{\pi}}e^{-X^{2}}\biggr\rbrack,
\label{it2}
\end{equation}
where
\begin{equation}
{\rm erf}(X)={2\over\sqrt{\pi}}\int_{0}^{X}e^{-t^{2}}dt,
\label{it3}
\end{equation}
is the error function, we find after some elementary calculations
that
\begin{equation}
D_{||}=2A{\rho} G(X){1\over v}, \label{it4}
\end{equation}
\begin{equation}
D_{\perp}=2A{\rho}\lbrack {\rm erf}(X)-G(X)\rbrack {1\over v},
\label{it5}
\end{equation}
\begin{equation}
{\mb\eta}=2A{\rho}\beta G(X){{\bf v}\over v}. \label{it6}
\end{equation}
These results are well-known \cite{bt} and are recalled here for
sake of completeness. We note that $D_{||}\sim v^{-3}$ for $|{\bf
v}|\rightarrow +\infty$. On the other hand, combining Eqs.
(\ref{tb10}), (\ref{it6}) and (\ref{it4}) we note that the
dynamical friction can be written in the form
\begin{equation}
\langle {\bf F}\rangle_{friction}=-2{\mb\eta}=-2\beta D_{||}{\bf v},
\label{it7}
\end{equation}
so that the  Einstein relation reads $\xi=2\beta D_{||}$ (the
factor $2$ is due to the velocity dependance of the diffusion
coefficient). Quite generally, in the thermal bath approximation,
we have
\begin{equation}
{\mb\eta}=\beta D_{||}{\bf v},
\label{it8}
\end{equation}
which results from Eqs. (\ref{tb16}) and (\ref{diff3}).

\subsection{Fermi-Dirac distributions}
\label{sec_fd}

We now consider the case of quantum particles (fermions) described
by the Fermi-Dirac entropy. In the thermal bath approximation, the
field particles are at statistical equilibrium with the
Fermi-Dirac distribution
\begin{equation}
f_{eq}({\bf v}_{1})={\eta_0\over 1+\lambda e^{\beta
{{v_{1}}^{2}\over 2}}}. \label{fd1}
\end{equation}
The parameter $\lambda$ is related to the density $\rho$ by
\begin{equation}
\rho={4\pi\sqrt{2}\eta_0\over \beta^{3/2}}I_{1/2}(\lambda),
\label{fd2}
\end{equation}
where
\begin{equation}
I_{n}(t)=\int_{0}^{+\infty}{x^{n}\over 1+te^{x}}dx,
\label{fd3}
\end{equation}
is the Fermi integral of order $n$. For the Fermi-Dirac entropy,
the functions $g$ and $H$ are given by $g(f)=f(1-f/\eta_0)$ and
$H(f)=f$. The diffusion coefficient and the friction coefficient
can be calculated explicitly by substituting Eq. (\ref{fd1}) in
Eqs. (\ref{Dperp2}), (\ref{Dpar2}) and (\ref{Fric2}), and
performing the integrals. Introducing the notation $X=\sqrt{\beta
v^{2}/ 2}$ and the incomplete Fermi integral
\begin{equation}
I_{n}(t,X)={1\over I_{n}(t)}\int_{0}^{X^{2}}{x^{n}\over 1+te^{x}}dx,
\label{fd4}
\end{equation}
we find after elementary calculations that
\begin{equation}
D_{||}=A {\rho}{1\over X^{2}}I_{1/2}(\lambda,X){1\over v},
\label{fd5}
\end{equation}
\begin{equation}
D_{\perp}=A {\rho}\biggl \lbrack {I_{-1/2}(\lambda)\over
I_{1/2}(\lambda)}I_{-1/2}(\lambda,X)-{1\over
X^{2}}I_{1/2}(\lambda,X)\biggr \rbrack {1\over v}, \label{fd6}
\end{equation}
\begin{equation}
{\mb\eta}=A {\rho}{\beta \over X^{2}}I_{1/2}(\lambda,X){{\bf
v}\over v}=\beta D_{||}{\bf v}. \label{fd7}
\end{equation}
These equations were previously derived in \cite{mnras} in a less neat
form. For $t\rightarrow +\infty$, we have the equivalent
\begin{equation}
I_{n}(t)\sim {1\over t}\Gamma(n+1).
\label{fd8}
\end{equation}
On the other hand,
\begin{equation}
I_{n}(+\infty,X)={2\over\Gamma(n+1)}\int_{0}^{X}y^{2n+1}e^{-y^{2}}dy.
\label{fd9}
\end{equation}
Therefore,
\begin{equation}
I_{1/2}(+\infty,X)=2X^{2}G(X), \qquad I_{-1/2}(+\infty,X)={\rm erf}(X).
\label{fd10}
\end{equation}
In the non-degenerate limit $\lambda\rightarrow +\infty$,
Eqs. (\ref{fd5})-(\ref{fd7}) return the results
(\ref{it4})-(\ref{it6}) valid for classical particles.

\subsection{Tsallis distributions}
\label{sec_ts}

We finally consider the case of generalized thermodynamics based
on Tsallis entropy. In the thermal bath approximation, the field
stars are at statistical equilibrium with the polytropic
distribution
\begin{equation}
f_{eq}({\bf v}_{1})=A\biggl (\lambda-{v_{1}^{2}\over 2}\biggr )^{1\over q-1},
\label{ts1}
\end{equation}
if $v\le v_{m}=\sqrt{2\lambda}$ and $f=0$ otherwise (we restrict ourselves to $q\ge 1$). The polytropic index $n$ is related to the $q$-parameter \cite{grand} by
\begin{equation}
n={3\over 2}+{1\over q-1}.
\label{ts2}
\end{equation}
The parameters $A$ and $\lambda$ are related to the generalized
temperature $\beta$ and to the density $\rho$ by the relations
\begin{equation}
A=\biggl\lbrack {(q-1)\beta\over q}\biggr \rbrack^{1\over q-1},
\qquad {\rho}=4\pi\sqrt{2} A\lambda^{n}B\biggl ({3\over
2},n-{1\over 2}\biggr ), \label{ts3}
\end{equation}
where
\begin{equation}
B(m,n)=\int_{0}^{1}x^{m-1}(1-x)^{n-1}dx,
\label{ts4}
\end{equation}
is the beta-function. For the Tsallis entropy, the $g$ and $H$
functions are $g(f)=f$ and $H(f)=f^{q}$. The diffusion coefficient
and the friction coefficient can be calculated explicitly by
substituting Eq. (\ref{ts1})  in Eqs. (\ref{Dperp2}),
(\ref{Dpar2}) and (\ref{Fric2}) and performing the integrals.
Introducing the notation  $X=\sqrt{(n+1)v^{2}/2\lambda}$ and the
incomplete beta-function
\begin{equation}
B_X(m,n)={1\over B(m,n)}{\int_{0}^{{X^{2}\over n+1}}x^{m-1}(1-x)^{n-1}dx},
\label{ts5}
\end{equation}
if $X\le \sqrt{n+1}$ and $I_X(m,n)=1$ otherwise, we find after
 elementary calculations that
\begin{equation}
D_{||}=A{\rho}{1\over X^{2}} B_{X}\biggl ({3\over 2},n+{1\over
2}\biggr ){1\over v}, \label{ts6}
\end{equation}
\begin{equation}
D_{\perp}=A{\rho} \biggl\lbrack 2 B_{X}\biggl ({1\over
2},n+{1\over 2}\biggr )-B_{X}\biggl ({3\over 2},n+{1\over 2}\biggr
){1\over X^{2}}\biggr\rbrack {1\over v}, \label{ts7}
\end{equation}
\begin{equation}
{\mb\eta}=A{\rho}{\beta \over X^{2}} B_{X}\biggl ({3\over
2},n+{1\over 2}\biggr ){{\bf v}\over v}=\beta D_{||}{\bf v}.
\label{ts8}
\end{equation}

For $n\rightarrow +\infty$, we have the equivalent
\begin{equation}
B(m,n)\sim {\Gamma(m)\over n^{m}}.
\label{ts9}
\end{equation}
On the other hand,
\begin{equation}
B_{X}(m,+\infty)={2\over\Gamma(m)}\int_{0}^{X}y^{2m-1}e^{-y^{2}}dy.
\label{ts10}
\end{equation}
Therefore,
\begin{equation}
B_{X}\biggl ({1\over 2},+\infty\biggr )={\rm erf}(X), \qquad B_{X}\biggl ({3\over 2},+\infty\biggr )=  2X^{2}G(X).
\label{ts11}
\end{equation}
In the limit $n\rightarrow +\infty$, corresponding to
$q\rightarrow 1$, Eqs. (\ref{ts6})-(\ref{ts8}) return the results
(\ref{it4})-(\ref{it6}) obtained in the context of ordinary
thermodynamics.

\section{Truncated models accounting for a permanent escape of particles}
\label{sec_escape}

In this section, we shall derive generalized truncated
distribution functions accounting for an escape of particles above
a limit energy $\epsilon_m$. For simplicity we shall work in the
thermal bath approximation. We thus describe the evolution of the
distribution function by the generalized Kramers equation
\begin{equation}
\label{esc1}
{\partial {f}\over \partial t}+{\bf  v}{\partial {f}\over\partial{\bf  r}}+{\bf F}{\partial {f}\over\partial{\bf  v}}={\partial\over\partial  {\bf v}}\biggl\lbrack {K\over v^{3}} \biggl (h(f){\partial {f}\over\partial   {\bf v}}+\beta g({f}) {\bf v}\biggr )\biggr\rbrack,
\end{equation}
where we have neglected anisotropy, taken
$D^{\mu\nu}=D_{||}\delta^{\mu\nu}$ in order to respect the
Einstein relation (\ref{it8}) and used the asymptotic expression
$D_{||}\sim v^{-3}$ of the diffusion coefficient valid for high
velocities. We assume that high energy particles are removed by a
tidal field (for globular clusters, this is the gravitational
attraction of a nearby galaxy). We seek therefore a stationary
solution of equation (\ref{esc1}) of the form $f=f(\epsilon)$
satisfying the boundary condition ${f}(\epsilon_{m})=0$, where
$\epsilon={v^2\over 2}+\Phi$ is the energy of a particle and
$\epsilon_{m}$ is the escape energy above which ${f}=0$. Using the
identity ${\partial\over\partial {\bf v}}({{\bf v}\over v^3})=0$
(valid for large $|{\bf v}|$), we obtain
\begin{equation}
\label{esc2}
{d\over d \epsilon}\Biggl\lbrack h(f) {d {f}\over d\epsilon}+\beta
g(f)\Biggr \rbrack=0,
\end{equation}
or, equivalently,
\begin{equation}
\label{esc3}
h(f){d {f}\over d \epsilon}+\beta g({f})=- J,
\end{equation}
where $J$ is a constant of integration representing physically a
current of diffusion. If $J=0$, the foregoing equation reduces to
\begin{equation}
\label{esc3bis}
C''(f){d {f}\over d \epsilon}+\beta =0,
\end{equation}
where we have used Eq. (\ref{ge5}). After integration, we recover the
usual equilibrium distribution
\begin{equation}
\label{esc3tris}
C'(f_{eq})=-\beta\epsilon-\alpha.
\end{equation}
If $J\neq 0$, Eq. (\ref{esc3}) accounts for an escape of
particles at a constant rate $J$. The system is therefore not
truly static since it looses gradually particles but we can
consider that it passes by a succession of quasi-stationary states
that are solution of Eq. (\ref{esc3}). Equation (\ref{esc3})
is a first order differential equation which can be integrated as
\begin{equation}
\label{esc4}
\int_{0}^{f}{h(t)dt\over g(t)+J/\beta}=\beta (\epsilon_{m}-\epsilon).
\end{equation}

Let us consider particular cases of this equation. For classical
particles described by the Boltzmann entropy, Eq. (\ref{esc4}) reduces
to
\begin{equation}
\label{esc5}
\int_{0}^{f}{dt\over t+J/\beta}=\beta (\epsilon_{m}-\epsilon).
\end{equation}
The integral is readily performed and we obtain the Michie-King
model
\begin{equation}
\label{esc6}
f=A (e^{-\beta\epsilon}-e^{-\beta\epsilon_{m}}),
\end{equation}
where we have set $A=(J/\beta)e^{\beta\epsilon_{m}}$. The Michie-King model
describes the tidally truncated structure of globular clusters in
astrophysics \cite{bt}. For quantum particles described by the
Fermi-Dirac entropy, Eq. (\ref{esc3}) reduces to
\begin{equation}
\label{esc7}
{d {f}\over d \epsilon}+\beta f(1-f/\eta_{0})=- J.
\end{equation}
This is a Riccatti equation that can be solved analytically. Assuming that
 degeneracy is
negligible for energies close to the escape energy, we get
\begin{equation}
{f}=\eta_{0}{e^{-\beta\epsilon}-e^{-\beta\epsilon_{m}}\over
\lambda+e^{-\beta\epsilon}}.
\label{esc8}
\end{equation}
This truncated distribution was previously derived in \cite{mnras}. It
could describe the case of galactic halos (e.g., massive neutrinos in
dark matter models) limited in extension by tidal forces. It could
also be of interest for collisionless stellar systems with
Lynden-Bell's interpretation of degeneracy. Finally, for systems of
particles described by the Tsallis entropy, Eq. (\ref{esc4}) reduces
to
\begin{equation}
\int_{0}^{\beta f/J}{t^{q-1}dx\over 1+t}={(J/\beta)^{1-q}\over q}\ \beta(\epsilon_{m}-\epsilon).
\label{esc9}
\end{equation}
For $q=1$, we recover Eq. (\ref{esc5}). If we introduce the function
\begin{equation}
\phi_{q}(x)=\int_{0}^{x}{t^{q-1}\over 1+t}dx,
\label{esc10}
\end{equation}
we can rewrite Eq. (\ref{esc9}) in the form
\begin{equation}
f={J\over\beta}\phi_{q}^{-1}\biggl\lbrack
{(J/\beta)^{1-q}\over q}\ \beta(\epsilon_{m}-\epsilon)\biggr\rbrack.
\label{esc11}
\end{equation}
For $q=1$, we recover the Michie-King model. A detailed study of these 
truncated models will be given elsewhere.

\section{Conclusion}
\label{conclusion}

In this paper, we have shown that standard kinetic equations
(Boltzmann, Landau, Kramers, Smoluchowski,...) can be generalized
so that they satisfy a $H$-theorem for an arbitrary functional of
the form $S=-\int C(f)d^{3}{\bf r}d^{3}{\bf v}$, where $C(f)$ is a
convex function. Boltzmann, Fermi, Bose and Tsallis entropies are
particular functionals of the above form. These generalized
kinetic equations have a thermodynamical structure which
corresponds either to a microcanonical (Boltzmann, Landau) or a
canonical (Kramers, Smoluchowski) description. One important
conclusion of our work is that Tsallis entropy does not play any
special role in this generalized thermodynamical formalism except
that leading to simple distributions \cite{gt}. Therefore, the
question that naturally emerges is whether other arguments can
give Tsallis entropy a fundamental justification or whether
Tsallis entropy is just a {\it simple} functional (associated to
power-laws) extending Boltzmann entropy and providing a good {\it
fit} of several observed phenomena. This is clearly an important
point to be settled in the future.

There are at least two distinct notions of generalized
thermodynamics. Generalized thermodynamics and kinetic equations
can arise in complex media when the transition probabilities have
an expression different from the one we would naively expect
\cite{kaniadakis}. We believe that generalized kinetic equations
are essentially {\it effective} equations attempting to take into
account ``hidden constraints'' that are not directly accessible to
the observer or that are difficult to formalize. If this idea is
correct, it means that it will never be possible to justify these
equations from first principle (except in toy models). This also
explains naturally why there is some indetermination in the theory,
either in Tsallis $q$-parameter or more generally in the function
$C(f)$. The important point in that context is to understand {\it
why} standard kinetic theory breaks up. To our point of view, it
is not sufficient to say ``since the system is non-extensive,
Boltzmann entropy is not correct and Tsallis entropy must be used
instead''. This argument is too simplistic and does not bring any
physical insight in the problem. The main interest of Tsallis
nonextensive thermodynamics (in addition to the development of a
consistent formalism) is to show that the naive Boltzmann description
fails in many systems. However, the reason of this failure has to
be understood in each case and this is the main challenge.

A notion of generalized thermodynamics also emerges in the context
of the violent relaxation of stellar systems and 2D vortices (or
other Hamiltonian systems with long-range interactions) described
by Vlasov-type equations. In that context, we can explicitly
illustrate the notion of ``hidden constraints'' that we mentioned
previously. According to rigorous statistical mechanics
\cite{lb,mr}, the metaequilibrium state resulting from complete
violent relaxation is obtained by maximizing the Boltzmann entropy
$S[\rho]$ for the fine-grained distribution $\rho({\bf r},{\bf
v},\eta)$ while conserving mass, energy and an infinity of additional
constraints played by the Casimirs. It turns out that the
coarse-grained distribution function $\overline{f}({\bf r},{\bf v})$
also maximizes a functional $S[\overline{f}]=-\int C(\overline{f})
d^{3}{\bf r}d^{3}{\bf v}$ while conserving only mass and energy
(robust constraints) \cite{grand,gt}. This functional, which could be
called a generalized entropy, is {\it non-universal} due to
fine-grained constraints that depend on the initial conditions.  The
Casimirs represent ``hidden constraints'' because, in practice, we
just know the coarse-grained field and we do not have access to the
initial conditions. In case of complete violent relaxation
$S[\overline{f}]$ is {\it never} Tsallis entropy since
$\overline{f}>0$ according to Lynden-Bell's theory. However, violent
relaxation is incomplete in general. The only thing we know for sure
is that the metaequilibrium state reached by the system is a
nonlinearly dynamically stable stationary solution of the Vlasov
equation on a coarse-grained scale. This implies in many cases that it
maximizes a $H$-function $S[\overline{f}]=-\int C(\overline{f})
d^{3}{\bf r}d^{3}{\bf v}$ at fixed mass and energy
\cite{grand}. There, $C(\overline{f})$ depends on the initial
conditions {\it and} on the efficiency of mixing. Tsallis entropy is a
particular H-function.  Since the dynamical stability criterion is
{\it similar} to a generalized thermodynamical criterion, we believe
that it is this {\it thermodynamical analogy} that justifies the
consideration of Tsallis functionals (they are not true entropies!) in
2D turbulence and stellar dynamics. However, we stress that Tsallis
entropy has no fundamental justification in that context \cite{brands}
and that, indeed, most vortices and galaxies are not polytropic
\cite{grand,gt}. Finally, we emphasize that the true statistical 
equilibrium state resulting from encounters between stars or between
point vortices (collisional relaxation) is described, in a suitable
thermodynamic limit, by the ordinary Boltzmann entropy although the
system is non-extensive and non additive. The pecularities due to the
absence of entropy maximum in gravitational systems correspond to
important physical processes (evaporation and gravothermal
catastrophe) and not to a break up of thermodynamics \cite{grand}. We
hope that this critical discussion will help to clarify the different
notions of ``generalized thermodynamis'' that appeared in the recent
literature.

\vskip1cm

I acknowledge stimulating discussions with F. Bouchet, T. Dauxois
and C. Sire.


\begin{thebibliography}{7}
%
\addcontentsline{toc}{section}{References}

\bibitem{balescu}  {\small R. Balescu, Statistical Mechanics of Charged Particles, Interscience, New York (1963).}

\bibitem{risken}  {\small H. Risken, {The Fokker-Planck equation} (Springer, 1989).}

\bibitem{gt}  {\small P.H. Chavanis, cond-mat/0209096. }

\bibitem{ipser}  {\small J.R. Ipser and G. Horwitz, ApJ
{\bf 232}, 863 (1979).}

\bibitem{tremaine}  {\small S. Tremaine, M. H\'enon and D. Lynden-Bell,  Mon. Not. R. astr. Soc. {\bf 219} 285 (1986).}

\bibitem{ellis}  {\small R. Ellis, K. Haven and B. Turkington, Nonlinearity {\bf 15}, 239 (2002).}

\bibitem{grand}  {\small P.H. Chavanis, A\&A {\bf 401}, 15 (2003).}

\bibitem{csr}  {\small P.H. Chavanis, J. Sommeria and
R. Robert, Astrophys. J. {\bf 471}, 385 (1996).}

\bibitem{rsmepp}  {\small R. Robert and J. Sommeria, Phys. Rev. Lett. {\bf 69}, 2776 (1992).}

\bibitem{kaniadakis}  {\small G. Kaniadakis, Physica A {\bf 296}, 405 (2001).}

\bibitem{bt}  {\small J. Binney and S. Tremaine, Galactic Dynamics, Princeton Ser. in Astrophysics, (1987).}

\bibitem{kin}  {\small P.H. Chavanis, Phys. Rev. E {\bf 58}, R1199
(1998); P.H. Chavanis \& C. Sire, Phys. Rev. E {\bf 62}, 490
(2001); P.H. Chavanis, Phys. Rev. E {\bf 64}, 026309 (2001).}

\bibitem{houches}  {\small  P.H. Chavanis, in {\it Dynamics and thermodynamics of systems
with long range interactions}, edited by Dauxois, T, Ruffo, S.,
Arimondo, E. and Wilkens, M. Lecture Notes in Physics, Springer
(2002) [cond-mat/0212223].}

\bibitem{lee}  {\small E.P. Lee, ApJ, {\bf 151}, 687 (1968).}

\bibitem{kandrupR}  {\small H.E. Kandrup, Physics Reports, {\bf 63}, 1 (1980).}

\bibitem{kandrup}  {\small H.E. Kandrup, ApJ, {\bf 244}, 316 (1981).}

\bibitem{dubrovnik} {\small P.H. Chavanis, in Proceedings of the Conference on
Multiscale Problems in Science and Technology (Springer 2002) [astro-ph/0212205].}

\bibitem{lb}  {\small D. Lynden-Bell, Mon. Not. R. astr. Soc.
{\bf 136}, 101 (1967).}

\bibitem{mr}  {\small J. Michel and R. Robert, Commun. Math. Phys.
{\bf 159}, 195 (1994).}

\bibitem{sw}  {\small P.H. Chavanis and J. Sommeria,
 Phys. Rev. E {\bf 65}, 026302 (2002).}

\bibitem{kp}  {\small B.B. Kadomtsev and O.P. Pogutse,  Phys. Rev. Lett.
{\bf 25}, 1155 (1970).}

\bibitem{sl}  {\small G. Severne and M. Luwel, Astrophys. \& Space Sci.  {\bf 72}, 293 (1980).}

\bibitem{mnras}  {\small P.H. Chavanis,  Mon. Not. R. astr. Soc.  {\bf 300}, 981 (1998).}

\bibitem{prl}  {\small P.H. Chavanis, Phys. Rev. Lett. {\bf 84}, 5512 (2000).}

\bibitem{kazantsev}  {\small E. Kazantsev, J. Sommeria and J. Verron, J. Phys. Ocean. {\bf 28}, 1017
 (1998).}

\bibitem{leith}  {\small C.E. Leith, Phys. Fluids {\bf 27}, 1388
 (1984).}

\bibitem{staquet}  {\small J. Sommeria, C. Staquet and  R. Robert, J. Fluid. Mech. {\bf 233}, 661 (1991).}

\bibitem{jm}  {\small G. Joyce and  D. Montgomery, J. Plasma Phys. {\bf 10}, 107 (1973).}

\bibitem{bs}  {\small F. Bouchet and  J. Sommeria, J. Fluid. Mech. {\bf 464},
165 (2002).}

\bibitem{montgomery}  {\small D. Montgomery, W.H. Matthaeus, W.T. Stribling, D. Martinez and S. Oughton,  Phys. Fluids A, {\bf 4}, 3 (1992).}

\bibitem{boghosian}  {\small B.M. Boghosian, Phys. Rev. E {\bf 53},
4754 (1996).}

\bibitem{brands}  {\small H. Brands, P.H. Chavanis, R. Pasmanter  and  J. Sommeria, Phys. Fluids {\bf 11}, 3465 (1999).}

\bibitem{rr}  {\small R. Robert and C. Rosier, J. Stat. Phys. {\bf 86}, 481 (1997).}

\bibitem{borland}  {\small L. Borland, Phys. Rev. E
{\bf 57}, 6634 (1998).}

\bibitem{sp}  {\small P.H. Chavanis, C. Rosier and C. Sire, Phys. Rev. E
{\bf 66}, 036105 (2002); C. Sire and P.H. Chavanis, Phys. Rev. E
{\bf 66}, 046133.}

\bibitem{gsp}  {\small P.H. Chavanis and C. Sire, cond-mat/0303088;
P.H. Chavanis, M. Ribot and C. Rosier, in preparation.}

\bibitem{biler}  {\small P. Biler and T. Nadzieja, preprint (2002).}



\end{thebibliography}
\end{document}